\pdfoutput=1
\documentclass[brainsci, %arts,
               review,submit,pdftex,moreauthors
               ]{Definitions/mdpi} 
\firstpage{1} 
\pubvolume{1}
\issuenum{1}
\articlenumber{0}
\pubyear{2022}
\copyrightyear{2022}
\datereceived{} 
\dateaccepted{} 
\datepublished{} 
\hreflink{https://doi.org/} % If needed use \linebreak
\usepackage{siunitx}%The siunitx package provides a  set  of  tools  for  authors  to  typeset  quantities  in  aconsistent  way.

\newcommand{\ms}{\si{\milli\second}}%

\usepackage{soulutf8}
\usepackage{color}

\usepackage[finalnew]{trackchanges}
\addeditor{Antoine}
\addeditor{Amélie}
\addeditor{Camille}
\addeditor{Jean}
\addeditor{JN}
\addeditor{Laurent}

\Title{Precise spiking motifs in neurobiological and neuromorphic data}
\TitleCitation{Precise spiking motifs}
\Author{
Antoine Grimaldi $^{1}$% $^{1,\dagger,\ddagger}$
\orcidA{}, 
Amélie Gruel $^{2}$\orcidB{}, % $^{2,\dagger,\ddagger}$,
Camille Besnainou $^{1}$\orcidC{},
Jean-Nicolas Jérémie $^{1}$\orcidD{},
Jean Martinet $^{2}$\orcidE{},
and Laurent Perrinet $^{1,}$\orcidF{}*}
\AuthorNames{Antoine Grimaldi, Amélie Gruel, Camille Besnainou, Jean-Nicolas Jérémie, Jean Martinet and Laurent Perrinet}
\AuthorCitation{Grimaldi, A.; Gruel, A.; Besnainou, C.; Jérémie, J.-N.;Martinet, J.; and Perrinet, L.}
\address{%
$^{1}$ \quad INT UMR 7289, Aix Marseille Univ, CNRS; 27 Bd Jean Moulin, 13005 Marseille, France\\
$^{2}$ \quad SPARKS,  Côte d'Azur, CNRS, I3S; 2000 Rte des Lucioles, 06900 Sophia-Antipolis, France
}
\corres{Correspondence: laurent.perrinet@univ-amu.fr}
\abstract{
Why do neurons communicate through spikes? By definition, spikes are all-or-none neural events which occur at continuous times. In other words, spikes are on one side binary, existing or not without further details, and on the other can occur at any asynchronous time, without the need for a centralized clock. This stands in stark contrast to the analog representation of values and the discretized timing classically used in digital processing and at the base of modern-day neural networks. As neural systems almost systematically use this so-called event-based representation in the living world, a better understanding of this phenomenon remains a fundamental challenge in neurobiology in order to better interpret the profusion of recorded data. With the growing need for intelligent embedded systems, it also emerges as a new computing paradigm to enable the efficient operation of a new class of sensors and event-based computers, called neuromorphic, which could enable significant gains in computation time and energy consumption --- a major societal issue in the era of the digital economy and global warming. In this review paper, we provide evidence from biology, theory and engineering that the precise timing of spikes plays a crucial role in our understanding of the efficiency of neural networks.
}
\keyword{spikes; asynchronous computing; neurobiology; computational neuroscience; neuromorphic engineering; heterogeneous delays; spiking motifs; polychronization.)} 
\begin{document}
%%%%%%%%%%%%%%%%%%%%%%%%%%%%%%%%%%%%%%%%%%
%
%%%%%%%%%%%%%%%%%%%%%%%%%%%%%%%%%%%%%%%%%%
%%%%%%%%%%%%%%%%%%%%%%%%%%%%%%%%%%%%%%%%%%
%%%%%%%%%%%%%%%%%%%%%%%%%%%%%%%%%%%%%%%%%%
\section{Introduction: Importance of precise spike timings in the brain}\label{sec:time}
%%%%%%%%%%%%%%%%%%%%%%%%%%%%%%%%%%%%%%%%%%
%%%%%%%%%%%%%%%%%%%%%%%%%%%%%%%%%%%%%%%%%%
%
%%%%%%%%%%%%%%%%%%%%%%%%%%%%%%%%%%%%%%%%%%
\subsection{Is there a neural code?}
%%%%%%%%%%%%%%%%%%%%%%%%%%%%%%%%%%%%%%%%%%
% CONTEXT
Neural activity is directly influenced by our immediate environment and by internal states and is structured in order to generate motor actions. The efficiency of these actions is key for survival, which is the sole determinant in the light of natural selection. A central question of modern neuroscience is to better understand the essence of neural activity, as exemplified by the recordings observed in neurobiological experiments. One sometimes uses the expression ``decoding the neural code'', although this implies the existence of a code, i.e. an explicit representation \add[Antoine]{of cognitive processes} within the neural activity. Nevertheless, we will use this terminology in all generality to denote the existence of a structure in neural activity. In this respect, it is \change[Antoine]{known}{reasonable to declare} that neural activity may be related to specific measurable variables\change[Antoine]{, and since Galvani's}{. Since Galvani's} experiments, we know that an electrical activity applied on muscular nerves can cause the stretching of a frog's limb (for a review, see~\citep{piccolino_luigi_1997}). A central and well-studied way of communication between neurons is specific electrochemical events called action potentials, or spikes, which were first discovered \change[Antoine]{by Lord Adrian}{at the beginning of the $XX^{th}$ century}~\citep{adrian_impulses_1926}. In this study, the frequency at which these spikes are emitted has been shown to be roughly commensurate with the stretch of the frog's limb.
In the scope of this article, we focus mainly on these spikes\change[Antoine]{which are brief (about one millisecond) impulses that propagate along the axons of neurons. First, these have the particular property of being essentially binary in their amplitude, i.e. they are prototypical, ``all-or-none''}{in vertebrate systems. They can be described as brief (about one millisecond) and prototypical, i.e. ``all-or-none'', impulses that propagate along the axons of neurons.}
 \change[Antoine]{Typically, action potentials are received by a neuron from other afferent neurons on the arborized ``input'' dendrite, integrated in its soma, and possibly further propagated along its ``output'' axon to reach efferent neurons.}{Typically, a postsynaptic neuron receives incoming spikes from other afferent neurons on the arborized ``input'' dendrite. The integration of these spikes by the dendritic tree and the soma of the postsynaptic neuron results in the modification of its membrane potential that possibly leads to the emission of an action potential along its ``output'' axon to reach efferent neurons.} \add[Antoine]{Except notably in the retina where neurons communicate with graded potentials}~\citep{gouras_graded_1960}\add[Antoine]{, it is assumed that spike trains are the main component of the neural code.}
\remove[Antoine]{The frequency at which these spikes are emitted has been shown to be roughly commensurate with the stretch of the frog's limb(Adrian citation), and this variable (measured in spikes per second, or Hertz) is as of today considered to be the main constituent of the neural code.} \change[Antoine]{Indded, neurophysiologists typically}{Until recently, most neurophysiologists} use the temporal evolution of the firing rate (for instance, as computed as the average occurrence of spikes in small temporal windows of about $100~\ms$) in order to characterize the dynamical activity of neurons. This may be extended by computing different statistics on each neuron's sequence of spikes~\citep{perkel_neuronal_1967} but also the dependence across neurons~\citep{perkel_neuronal_1967-1}. 

% GAP
However, computational neuroscience models have suggested that the precise timing within a sequence of spikes may play a crucial role and that neurons \change[Antoine]{may be as well integrators than synchrony detectors}{may be synchrony detectors as well as integrators}~\citep{abeles_role_1982}. \add[Laurent]{In particular, it is possible that the minute arrangement of temporal delays between neurons may provide a computational advantage.} We will investigate this very hypothesis in this review. %Indeed, the response of one biological neuron largely depends on the precise timing of the sequence of incoming spikes that reach its soma~\citep{paugam-moisy_computing_2012}. 
In comparison to a classical \change[Antoine]{analogical}{analog} vector of inputs, this \emph{event-based representation} observed in the neural code is essential in understanding information processing~\citep{carr_processing_1993}. For instance, it expands the capabilities of representations \change[Antoine]{(which rely on the firing rate)}{of the rate coding hypothesis that relies only on the firing rate} by considering \change[Antoine]{a representation based on repetitions of spiking motifs at precise times of occurrence.}{representations based on the precise timing of single spikes.} \remove[Antoine]{This hypothesis is directly inspired by neurobiological observations and expands the capabilities of analog representations (which are based on spiking firing rates) by considering a representation based on instances of spiking motifs at precise times of occurrence.} Additionally, numerous studies demonstrated the importance of precise timing in neural population activity~\citep{davis_spontaneous_2021}, efficient encoding thanks to the use of spike latencies~\citep{perrinet_coding_2004,gollisch_rapid_2008} or precise timing in the auditory system~\citep{deweese_binary_2003,carr_circuit_1990}. All these findings, and more~\citep{bohte_evidence_2004,dilorenzo_spike_2013}, highlight the importance of the temporal aspect of the neural code and further suggest the existence of precise spatio-temporal spiking motifs in the input which excites neurons. A mathematical formalization would be particularly well suited to neuromorphic computing~\citep{roy_towards_2019}, and would allow for the supervised or self-supervised learning of such motifs in any event-driven data.  Crucially, validating this hypothesis would also be crucial in our understanding of neural processes. %
%
%%%%%%%%%%%%%%%%%%%%%%%%%%%%%%%%%%%%%%%%%%
\subsection{Dynamics of vision and consequences on the neural code}%Precise timing in the neural code enables efficient vision}
%%%%%%%%%%%%%%%%%%%%%%%%%%%%%%%%%%%%%%%%%%
%
Let us start with a focus on the state-of-the-art of the role of dynamics in vision. Broadly speaking, vision is the set of processes that allow us \change[Antoine]{to make sense of the luminous world}{to make sense of the world through luminous signals}, and is an intensively studied field in neuroscience, in particular with respect to deciphering the neural code. In most mammals, light enters the eye to induce neural activity on the retina, which maintains a certain similitude between the topology of external visual space and its representation on the retina, called retinotopy. The origins of this neuroscientific question can be found in the first experiments from Pierre Flourens which, using lesions in animals, demonstrated the relationship between visual sensations and activity in the cerebral cortex~\citep{flourens_recherches_1842,pearce_marie-jean-pierre_2009}. This was also observed \remove[Antoine]{by Lord Adrian} when recording the activity of the frog's visual system~\citep{adrian_impulses_1926}. In a series of seminal studies,~\citet{hubel_receptive_1968} showed that this activity could be selective to different features, such as the visual orientation or motion of elementary contours. For a large proportion of neurons, there is \change[Antoine]{remarkably a monotonic}{a remarkably monotonic} relationship between the contrast of visual features and the firing frequency of neurons. However, there is no consensus to explain the multiple nonlinear mechanisms that transform the visual scene into retinotopic neural activity maps, even though these processes seem to constitute essential pieces to this puzzle~\citep{carandini_normalization_2012}. 
%----------------------------%
\begin{figure}
\centering
\includegraphics[width=.7\textwidth]{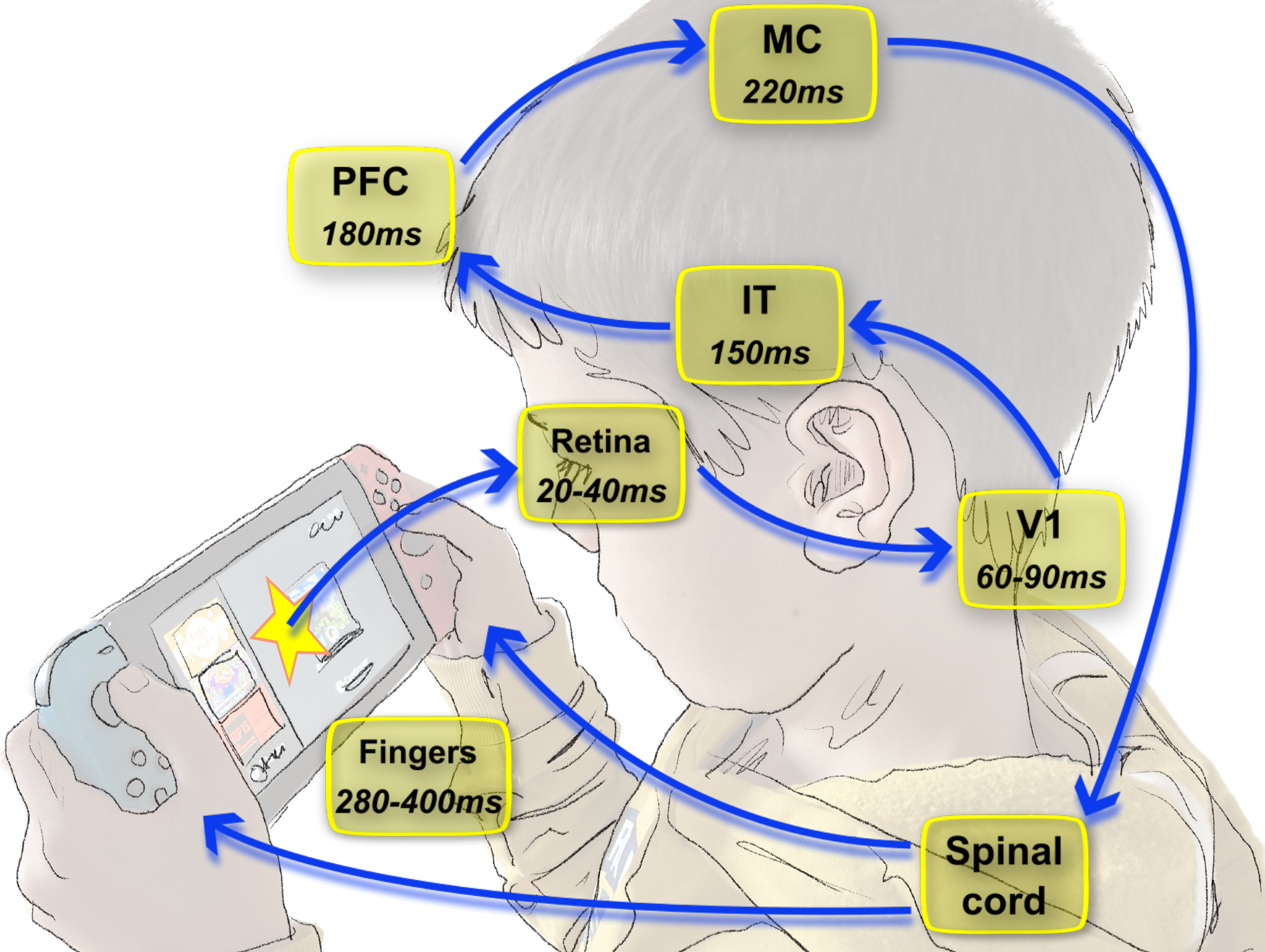}
% thorpe
% \includegraphics[width=.7\textwidth]{figures/visual-latency-estimate.jpg}%-estimate.jpg}
\caption{{\bf Latency of the different processing steps along the human visual pathway.} Though the visual system is highly inter-connected, one can follow the sequence of activations whenever an image (here a yellow star) is flashed in front of the eyes. Different areas are schematically represented by ellipses and arrows denote the fastest feed-forward activation, ordered with respect to their activation latency in $\ms$. In order, the retina is first activated ($20-40~\ms$), then the thalamus and the primary visual cortex (V1, $60-90~\ms$). This visual information  projects to the temporal lobe to reach the infero-temporal area (IT, $150~\ms$) for object recognition. It then reaches the prefrontal cortex (PFC, $180~\ms$) which modulates decision making to evoke the Motor Cortex (MC, $220~\ms$) which may mediate an action. This is eventually relayed through the spinal cord to trigger finger muscles, with latencies about $280-400~\ms$.}\label{fig:thorpe}
\end{figure}
%----------------------------%
In particular, there have been some remarkable findings when studying the dynamics of vision. For instance, Simon Thorpe's group has shown during the last decades numerous examples demonstrating that humans can categorize briefly presented images in a fraction of a second. \change[Antoine]{His}{Their} experiments consisted in asking subjects to categorize images that do or do not contain animals~\citep{thorpe_speed_1996}. The results showed that humans were able to perform this task very well (with a success rate of more than 95\%) but above all that a differential activity for the two categories of images could be observed by electroencephalography, showing that this differentiation emerges with a very short latency in neural activity. These results have been extended to several species, including primates. Different experimental protocols have shown for example that the motor response could be extremely fast (of the order of $120~\ms$) when the task was to perform a saccade~\citep{kirchner_ultra-rapid_2006}. This fast processing correlates with the surprising experiments of fast serial detection, which consists in presenting a fast succession of different images and decoding via the EEG if the observer can detect for example the presence of an animal~\citep{keysers_speed_2001}. As expected, the performances decrease progressively as the frequency of presentation of the images increases. However, it has been shown in the macaque that a significant performance could be maintained with an image presentation time of only $14~\ms$ per image.

Although surprising, this speed of the visual cortex in primates is compatible with the latencies that are recorded at the neuro-physiological level. The rapid propagation of the visual information in the thalamus, then in the primary visual cortex takes about $45~\ms$ in the macaque~\citep{schmolesky_signal_1998} and about $60~\ms$ in humans~\citep{vanni_coinciding_2001}. This functioning of visual processing as a forward pass is most prominent in fast processing (see Figure~\ref{fig:thorpe}), and can be complemented with feedback loops from the higher areas to the sensory areas~\citep{lamme_distinct_2000}. An important consequence of this speed of processing of vision is that it implies that processing is carried out using only very few spikes per layer. As a comparison, the latencies in macaque monkeys are about: Retina, $20-40~\ms$; V1, $40-60~\ms$; IT, $80-100~\ms$; MC, $140-190~\ms$; to finger muscles, $180-260~\ms$.\add[Laurent]{Note that, since maximal conduction speeds are roughly constant, theses latencies are comparable to that found in humans, with a ratio given by the physical size of the whole system.} It follows that if we consider that a behavioral response occurra in only $200~\ms$, it would involve about ten processing stages along the ``forward'' pathways of the visual system. Such processes were indeed efficiently reproduced in feed-forward models trained with back-propagation~\citep{serre_feedforward_2007,jeremie_ultrafast_2022}. At the same time, it has been demonstrated that one spike requires a significant amount of time (about $10~\ms$) to be conducted from one layer to the next~\citep{nowak_timing_1997,thorpe_seeking_2001}.  This figure is inspired by similar schematics done for monkeys in~\citep{thorpe_seeking_2001}. As a consequence, these results suggest that ``like other senses, vision relies heavily on temporal strategies and temporal neural codes to extract and represent spatial information''~\citep{rucci_temporal_2018}.
%%%%%%%%%%%%%%%%%%%%%%%%%%%%%%%%%%%%%%%%%%
\subsection{How precise spike timing may encode vectors of real values}
\begin{figure}
\centering
\includegraphics[width=\textwidth]{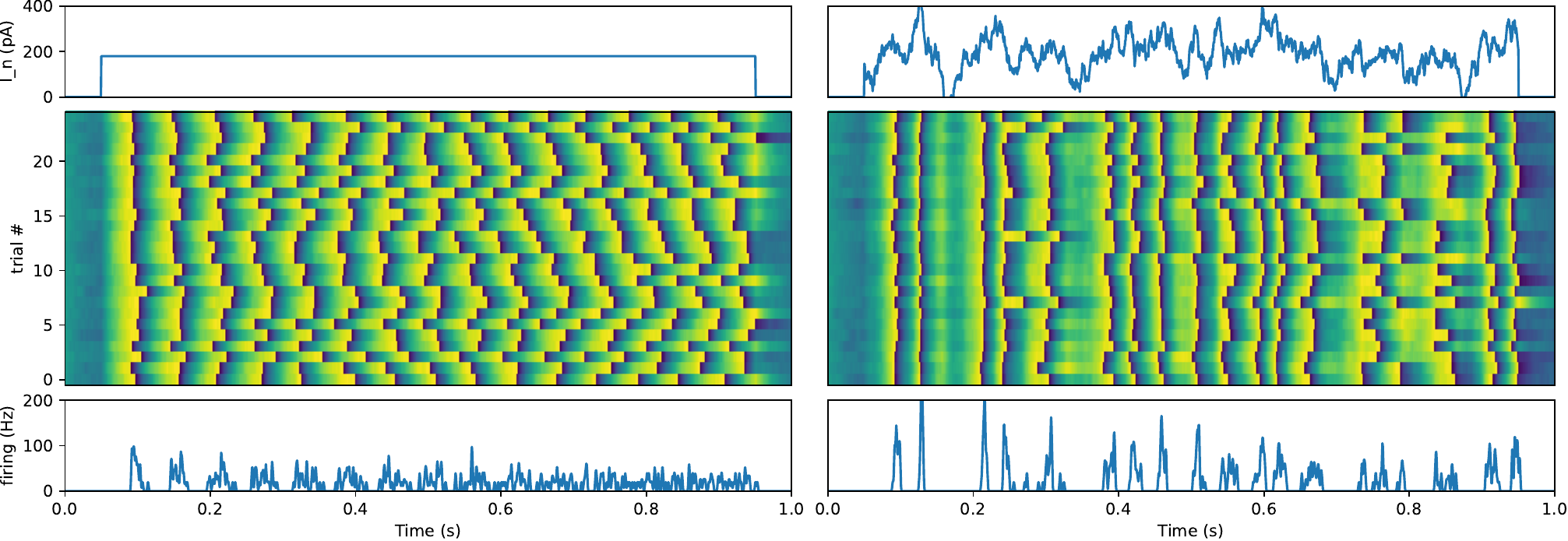} 
\caption{\textbf{Reproducibility of the spiking response of a neuron. }The timing of the spikes produced following the repetition of a step stimulus is less reproducible than that to a noisy stimulus. The stimulus current value over time for a step stimulus (top left) and for a noisy one (top right). Trial repetitions of a Leaky Integrate-and-Fire neuron stimulated by the stimulus on the upper row (middle row). Membrane potential is represented by dark blue color when light yellow colors when depolarized) and quantified by the average firing rate across trials (lower row). While this seems paradoxical at first sight, it highlights the consequence of using the same \emph{frozen} noise at each repetition and highlights the highly reproducible pattern of spikes when it is driven by a highly dynamic input. See this~\href{https://github.com/laurentperrinet/2022_UE-neurosciences-computationnelles/blob/master/C_MainenSejnowski1995_Perrinet.ipynb}{notebook} for a replication of the results from~\citep{mainen_reliability_1995} using a simple LIF model.}\label{fig:mainen}
\end{figure}

%%%%%%%%%%%%%%%%%%%%%%%%%%%%%%%%%%%%%%%%%%
\add[Laurent]{Let's now focus on one processing step along a cortical pathway. }
% reliability of spike timings
Sensory data is most often represented by continuous values, such as the energy produced by a flow of photons that hits the different photoreceptors of the retina. How may such information be encoded in neuronal activity? The analysis of generic raster plots reveals particular traits that hint at the role of a precise timing. For instance, the firing rate of cortical cells in awake monkeys is highly irregular~\citep{softky_highly_1993}, which makes it at first sight inconsistent with the temporal integration of firing rate. Remarkably, it has been observed that the response of a neuron in a cortical slice to a current step could be highly non-reproducible: while the first spike is aligned to stimulus' time, the subsequent spike times tend to diffuse for independent repetitions of the stimulation~\citep{bryant_spike_1976,mainen_reliability_1995} (see Figure~\ref{fig:mainen}). However, if that same neuron is now driven by a \emph{frozen} noise, that is, a highly dynamic signal which is repeatedly replayed from trial to trial, then the output spikes are highly reproducible (for a review, see~\citep{ermentrout_reliability_2008}). This is consistent with the differential role of different stimulus frequencies (for instance the gamma range around $80~\si{\Hz}$) on the reliability of spike timing reported in~\citep{nowak_influence_1997} : ``we found that, as expected given the resistive and capacitive properties of cortical neurons, low frequencies have a larger effect on the membrane potential of cortical neurons than do higher frequencies. However, increasing the amount of gamma range fluctuations in a stimulus leads to more precise timing of action potentials.'' 

% retina
\remove[Antoine]{How may information about the sensory world be encoded in the spiking pattern?} At the level of the retina, it has been shown that a coding of luminance values in the image using the timing of the spikes may be at work~\citep{gollisch_rapid_2008}. In particular, these results show that the response of ganglion cells to the visual gratings projected on the retina could be encoded in the latency of the response and not only in the frequency of the discharge, as it is often assumed. These results have been extended to natural images and show a qualitatively similar behavior. The authors' conclusion is that the precise spiking latency of the neurons encodes the spatial features of the image. Interestingly, such a precise latency mechanism may underlie some visual illusions, e.g., the false color illusion in the \href{https://michaelbach.de/ot/col-Benham/index.html}{Benham Top} based on center-surround interactions in the parvocellular pathway~\citep{kenyon_theory_2004}. This evidence found in the retina can be extended to other areas such as the visual cortex.
%----------------------------%
% C_MainenSejnowski1995_Perrinet.ipynb
%----------------------------%

% %----------------------------%
% \begin{figure}
% \centering
% \mynote{redo a figure similar to the thorpe code}%\includegraphics{images/roc.jpg}
% \caption{Latency coding. An input analog profile is encoded in latencies: the higher the contrast, the shorter the latency. In this example, one generates at most one spike per neuron.
% }\label{fig:roc}
% \end{figure}
% %----------------------------%
In fact, similar results have been demonstrated through neurophysiological recordings in the primary visual cortex and show that different levels of visual activity will induce different levels of neuronal discharge latency in the primary visual area~\citep{celebrini_dynamics_1993}. First-spike latency codes are a feasible mechanism for information transfer, even when biologically plausible estimates of stimulus onset are considered, for instance for sound localization~\citep{chase_first-spike_2007}.  Note also that timing is not entirely sensory or internal, but can be used as a general neural coding principle. In~\citep{safaie_turning_2020} for instance, they found that ``timing accuracy was improved when the environment afforded cues that rats can incorporate into motor routines. Timing, at least in animals, may thus be fundamentally embodied and situated.''  Many models have used these properties in temporal coding to build fast image categorization networks~\citep{gautrais_rate_1998,delorme_spikenet_1999, perrinet_coding_2004}. These models take the form of artificial spiking neural networks (SNNs) and have been able to demonstrate their practical applications for image categorization~\citep{delorme_ultra-rapid_2000}. One of these is the SpikeNet algorithm, which uses a purely temporal approach by encoding information using one spike per neuron by using the rank of neurons' activation~\citep{delorme_spikenet_1999,bonilla_analyzing_2022}. Another class of artificial SNNs uses precise spike timing as a metric in order to determine the structure of the network in order to minimize a cost function. This was implemented in the SpikeProp algorithm~\citep{bohte_error-backpropagation_2002} and has been extended in novel gradient-based methods. The subsequent surrogate gradient method is now widely used in methods that attempt to transfer performance from analog (CNNs) to spike-based (SNNs) architectures~\citep{zenke_remarkable_2021}. This type of modeling often uses the classical task of categorizing images developed in deep learning, while adapting it to the specificity of the event-based representation~\citep{goltz_fast_2021}. For instance,~\citep{kheradpisheh_stdp-based_2018} implements a STDP-based spiking deep convolutional neural networks for object recognition or~\citep{tavanaei_representation_2018} develops a form of spike-based, competitive learning applied for unsupervised learning. 
However, the performance of SNNs is still lagging compared to that of analog networks and the question of the advantage of using spikes in machine learning and computer vision remains open. \change[Antoine]{Achieving to efficiently implement this type of modeling is extremely important with respect to the development of a new generation of event-based cameras}{Improvements in this new generation of Artificial Neural networks (ANN) would bring major advances in terms of efficient computations in machine learning. They would benefit in particular to a new generation of cameras called Silicon Retinas}~\citep{gallego_event-based_2022} (see Section~\ref{sec:neuromorphic}). \remove{These represent visual information as events, similarly to the biologically-inspired representation described above and in contrast with a basic frame-based representation}%~\citep{lichtsteiner_128x128_2008} (for a review, see~\citep{rasetto_challenges_2022}, this will be developed in Section~\ref{sec:neuromorphic}).This work has been extended to include unsupervised learning capabilities, and we have recently developed a SNN architecture that allows to categorize images of different classes in only a few spikes~\citep{grimaldi_homeostatic_2021,grimaldi_robust_2022}. 

\change[Antoine]{To alleviate this gap in computational neuroscience, we will further review the role of  precise spike timing in neural assemblies from a neuroscientific perspective in the next section.}{Even if technology lags far behind biology, this introduction demonstrates the importance of timing in neural processes and we will further review the role of precise spike timing in neural assemblies. We start by reviewing the different hypotheses that aim at deciphering the neural code with spatio-temporal spiking motifs. After listing some biological evidence for the use of precise spike timing we review some computational models and neuromorphic technics that add this temporal dimension to their computations.}
%
%%%%%%%%%%%%%%%%%%%%%%%%%%%%%%%%%%%%%%%%%%%%%%%
\section{Role of precise spike timing in neural assemblies}
%%%%%%%%%%%%%%%%%%%%%%%%%%%%%%%%%%%%%%%%%%%%%%%
%\add[Antoine]{In this first section, we have introduced the notion of rate coding and demonstrated that spike timing can also carry information. There are multiple experimental evidences that neurons use different strategies to communicate.}~\cite{maunsell_functional_1983} \add[Antoine]{demonstrate that neurons of visual area MT in macaque monkeys encode motion direction through population activity, offering robustness to single neuron activity fluctuations. For rapid image processing, it has been observed that neurons in the salamander retina represent the spatial structure of an image with the relative timing of their first spikes}~\citep{gollisch_rapid_2008}\add[Antoine]{. In the primary visual area of macaque monkeys, the phase of firing convey information about naturalistic stimuli that cannot be described by firing rates alone}~\citep{montemurro_phase-of-firing_2008}\add[Antoine]{. Among a variety of strategies to communicate, so far we have also evidence that neurons can use correlation coding}~\citep{decharms_primary_1996}\add[Antoine]{ or sparse coding}~\citep{vinje_sparse_2000}. 
\add[Antoine]{In this first section, we have introduced the notion of rate coding and demonstrated that spike timing can also carry information. Scientists found experimental evidence for various hypotheses of neural representations like population coding}~\citep{maunsell_functional_1983}\add[Antoine]{, time-to-first-spike coding}~\citep{gollisch_rapid_2008}\add[Antoine]{, phase-of-firing coding}~\citep{montemurro_phase--firing_2008}\add[Antoine]{, correlation coding}~\citep{decharms_primary_1996}\add[Antoine]{ or sparse coding}~\citep{vinje_sparse_2000}\add[Antoine]{. In the scope of this review, we infer that spike trains are composed of repeating spiking motifs and we focus on precise spatio-temporal representations composed of a motif of spikes defined precisely in time and in the presynaptic address space. In all generality, this representation can encompass all the previous ones except for rate coding which is not defined locally in time. In this section, we choose to described different hypotheses making use of spatio-temporal patterns of spikes that can be propagated among neural assemblies.}
%%%%%%%%%%%%%%%%%%%%%%%%%%%%%%%%%%%%%%%%%%%%%%%
\subsection{One first hypothesis: synchronous firing in cell assemblies}
%%%%%%%%%%%%%%%%%%%%%%%%%%%%%%%%%%%%%%%%%%%%%%%
%
In his book ``Corticonics'',~\citet{abeles_corticonics_1991} queried whether the role of cortical neurons is to integrate synaptic inputs or rather to detect coincidences in temporal spiking patterns. The book gradually leads the reader from the macroscopic cortical anatomy and standard electro-physiological properties of single neurons to neural network models. While the first hypothesis favors the rate coding theory, the second possibility highlights the need for temporal precision in the neural code~\citep{abeles_role_1982,paugam-moisy_computing_2012}. The book then demonstrates that neural assemblies could form so-called ``synfire chains,'' that is, showing the emergence of synchronous activity on subsets of neurons which could be propagated in a stable fashion. More broadly, the idea of using the synchrony of co-activation in a cell assembly is reminiscent of the hypothesis that was formalized by~\citet{hebb_organization_1949}: \add[Antoine]{"cells that fire together wire together"}. Since this date, multiple experimental observations have suggested the existence of this precise zero-phase-lag spike synchronization in a defined subset of neurons~\citep{harris_organization_2003}. One possible function of this synchronization may serve the binding of information distributed in the brain~\citep{singer_visual_1995, roelfsema_visuomotor_1997}.

Some experimental results show the emergence of synchrony, for instance in motor cortical function~\citep{riehle_spike_1997}. Interestingly, these authors showed that ``accurate spike synchronization occurred in relation to external events (stimuli, movements) and was commonly accompanied by discharge rate modulations but without precise time locking of the spikes to these external events. Spike synchronization also occurred in relation to purely internal events (stimulus expectancy), where firing rate modulations were distinctly absent. These findings indicate that internally generated synchronization of individual spike discharges may subserve the cortical organization of cognitive motor processes.'' Moreover, such emergence could change over the learning period involved in learning a task~\citep{kilavik_long-term_2009} and showed some tuning to movement direction and reaction time~\citep{grammont_spike_2003}. It is important to note that synchronous events tend to lock to spatio-temporal patterns of neural activity called \href{https://en.wikipedia.org/wiki/Local_field_potential}{LFP} beta waves~\citep{denker_lfp_2018} and were also extended to larger assemblies using statistical methods~\citep{torre_asset_2016} (see Section~\ref{sec:detection} for further details). \change[Antoine]{Note also that synchronicity may explain some unintuitive results. Indeed, it has been shown that thalamo-cortical synapses are relatively weak compared to the amount of intracortical activity. However, this pathway is sufficient to drive the cortex, as this input is more often synchronously active, allowing it to more efficiently drive cortical neurons}{Synchronicity is also an interesting proposition to explain how the relatively weak thalamo-cortical synapses are able to drive cortical neurons. Among different explanations including travelling waves}~\citep{ben-yishai_traveling_1997}\add[Antoine]{, a synchronous activity at the synaptic level may be sufficient to elicit activity in the cortex.}~\citep{bruno_cortex_2006}. 

\begin{figure}
\centering
\includegraphics[width=\textwidth]{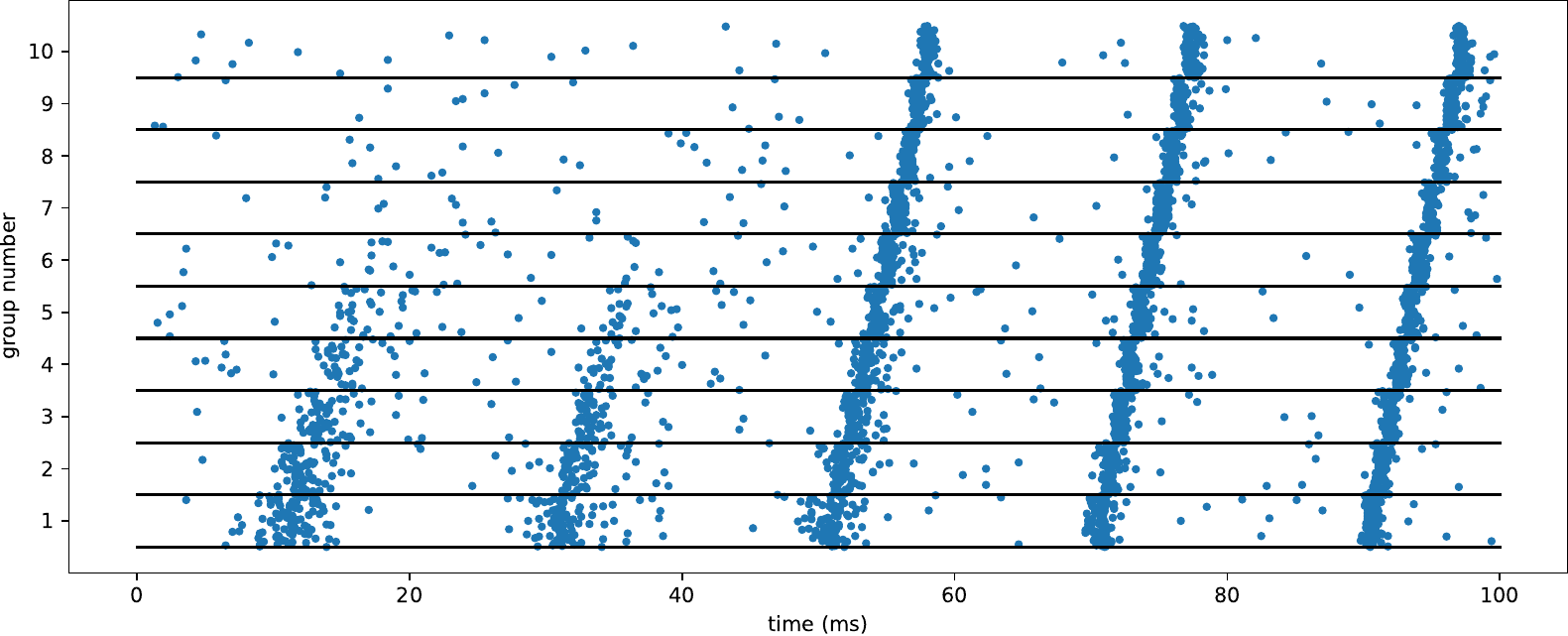}
\caption{\textbf{Simulation of a synfire propagation \href{https://brian2.readthedocs.io/en/stable/examples/frompapers.Diesmann_et_al_1999.html}{using Brian}.} The model consists of $10$ groups (arranged with the first group represented in the lowest row) of $100$ neurons each. Five pulses with decreasing jitter are generated in the first group around times $10$, $30$, $50$, $70$ and $90$~\ms~(with jitters given by a standard deviation which linearly decreases from $5$ to $1$~\ms). This generates a pulse after a certain processing delay in the second group with a different jitter. While the first two pulses progressively vanish in the following groups, starting from the third input pulse (with a jitter of $3~\ms$), it is propagated to the following groups. This allows the propagation of the synchronous activity along the chain of the neural groups.}\label{fig:diesman}
\end{figure}

Theoretically\add[Antoine]{, even if vertebrate's neural system is not likely to be modeled only by such deterministic connectivity}~\citep{deneve_bayesian_2004, ballard_dual_2011}, it was shown that a simple model may allow the propagation of such synfire chains~\citep{gewaltig_propagation_2001} by considering the dynamics of leaky integrate-and-fire (LIF) neurons in different groups of similar size. Each neuron of one group is connected by an excitatory synapse to the next. When a pulse is elicited in the first group, this may generate a spike in the next group. Depending on the concentration of synaptic weight values, this new activity may become more or less synchronized with respect to that of the previous pulse (as measured by the standard deviation of spike times within the pulse). Recursively applying this to a sequence of groups within a chain generates either a synfire propagation or not. A simple simulation of synfire propagation is shown in Figure~\ref{fig:diesman}. A crucial aspect of this emergence is explained by the dynamics of the spiking neuron model~\citep{gerstner_time_1995} and in particular the balance between excitation and inhibition~\citep{azouz_stimulus-selective_2008}. This balance was for instance modeled by feed-forward inhibition, a fine-scaled latency mechanism that is an essential ingredient in modelling so-called push-pull effects in the primary visual cortex~\citep{kremkow_push-pull_2016}. Further models have shown that such synfire chains could be embedded in topographies~\citep{aviel_embedding_2003}, while others used conductance-based neurons with feed-forward inhibition to improve the robustness of the propagation~\citep{kremkow_functional_2010}. In particular, this was implemented as a computational neuroscience benchmark model using the pyNN language~\citep{davison_pynn_2008} both in CPU-based and neuromorphic hardware~\citep{pfeil_six_2013}. % (see Figure~\ref{fig:pynn}).
%
%----------------------------%
%\begin{figure}
%\centering
%\includegraphics[width=.7\textwidth]{figures/fnins-07-00011-g004.jpg}
%\caption{A pyNN implementation of a Synfire chain with feedforward inhibition. The background noise  is implemented as random Gaussian current. The state space gives %the duration of synfire chains as a function of the protocol's parameters.}\label{fig:pynn}
%\end{figure}
%----------------------------%
%----------------------------%
%----------------------------%

Attempts have been made to detect such synfire chains in neurobiological data.~\citet{schrader_detecting_2008} envisioned that ``sensitivity is high enough to detect synfire chain activity in simultaneous single-unit recordings of 100 to 200 neurons from such data, enabling application to experimental data in the near future.'' Indeed, simultaneously recorded activities of neurons in the primary motor cortex of monkeys exhibited context-dependent, rapid changes in the patterns of coincident action potentials~\citep{riehle_spike_1997,grammont_precise_1999}. It is now commonly accepted that planning and execution of movements are based on distributed processing by neural populations in motor cortical areas yet it is less clear \remove[Antoine]{, though,} how these populations organize dynamically to cope with the momentary computational demands.  In~\citep{brette_computing_2012}, the author proposes a simple spike-based computational framework, based on the idea that stimulus-induced synchrony can be used to extract sensory invariants (for example, the location of a sound source), which is a difficult task for classical ANNs. It relies on the simple remark that a serie of repeated coincidences is in itself an invariant. Many aspects of perception rely on extracting invariant features, such as the spatial location of a time-varying sound, the identity of an odor with fluctuating intensity, or the pitch of a musical note.

This is also expressed in the idea that different cortical areas could achieve binding by synchrony~\citep{fries_mechanism_2005}. The synchronicity will generate rhythms at different ranges of frequencies, with spikes arriving at peak susceptibility (top of a cycle) or down. Such a theory has surprisingly been validated in EEG recordings to explain for example the continuous wagon-wheel illusion, i.e. the perceived reversal of the rotational movements of the spikes in a rotating wheel~\citep{vanrullen_continuous_2006}. More generally, it can be shown that the phase of alpha oscillations (about $10$ Hz) is causally linked with modulations of cortical excitability and with visual perception~\citep{dugue_phase_2011}. The question still remains open as to know if this is an epiphenomenon or a working mechanism of the neural code.
\subsection{A further hypothesis: travelling waves}
To further investigate the role of precise timing, let us also focus on the role of differential timings in an assembly of neurons. As we have seen, a visual feature will induce the firing of different cells at different latencies~\citep{celebrini_dynamics_1993}. Using intracellular recordings, it was shown that the response to a focal visual activation would elicit a latency basin, that is, a graded onset of the neural response from the most activated to neighboring neurons~\citep{bringuier_horizontal_1999}. In particular, it was shown that the network of the so-called horizontal connections within a cortical area is typically not myelinated and that this latency basin would be determined by the propagation speed within that area. For generic visual scenes, these processes would generate a complex interplay between the dynamics of the sensory signal and the spatio-temporal determinants of these interactions. Such interactions may underlay anticipatory mechanisms in the primary visual cortex~\citep{benvenuti_anticipatory_2020,le_bec_horizontal_2022}. The underlying process could be the emergence of propagating waves on the surface of the cortex.

Propagating waves in the neuronal response occur in many excitable media and were found in neural systems, for instance in the retina\note[Antoine]{added ref to Feller}~\citep{feller_dynamic_1997} or the neocortex~\citep{bienenstock_model_1995}. While propagating waves are clearly present under anesthesia, whether they also appear during awake and conscious states remained unclear until recent discoveries. One possibility is that these waves were systematically missed in trial-averaged data, due to variability. A recent work~\citep{muller_stimulus-evoked_2014} presents a method for detecting propagating waves in noisy multichannel recordings. Applying this method to single-trial voltage-sensitive dye imaging data\note{The reviewer is right to point that calcium imaging may not be the best tool to capture precise spiking motif as both spatial and temporal resolution are not that great. We argue that few imaging techniques can capture the precise spike location, even good spike sorting algorithms. Also, here we study a propagation of activity defining a spatio-temporal pattern that may not have the same precision as some spiking motifs we describe later.}, the authors show that the stimulus-evoked population response in the primary visual cortex of the awake monkey propagates as a travelling wave, with consistent dynamics across trials. A network model suggests that this reliability is the hallmark of the horizontal fiber network of superficial cortical layers. Propagating waves with similar properties occur independently in the secondary visual cortex, but maintain precise phase relations with the waves in the primary visual cortex. These results show that, in response to a visual stimulus, propagating waves are systematically evoked in several visual areas, generating a consistent spatio-temporal frame for further neural interactions.

More recently, novel multi-unit recording techniques have enabled the identification of travelling waves of neural activity in different areas of the cortex~\citep{muller_cortical_2018}. Authors review these findings by considering the mechanisms by which travelling waves are generated, and evaluate their possible roles in cortical function. In particular, spontaneous travelling waves naturally emerge from horizontal fiber time delays and travel through locally asynchronous-irregular states~\citep{davis_spontaneous_2021}. Studies of sensory-evoked neuronal responses often focus on mean spike rates, with fluctuations treated as internally-generated noise. However, fluctuations of spontaneous activity, often organized as travelling waves, shape stimulus-evoked responses and perceptual sensitivity. The mechanisms underlying these waves are unknown. Further, it is unclear whether waves are consistent with the low rate and weakly correlated ``asynchronous-irregular'' dynamics observed in cortical recordings. In that paper, the authors describe a large-scale computational model with topographically-organized connectivity and conduction delays relevant to biological scales. They find that spontaneous travelling waves are a general property of these networks. The travelling waves that occur in the model are sparse, with only a small fraction of neurons participating in any individual wave. Consequently, they do not induce measurable spike correlations and remain consistent with locally asynchronous irregular states. Further, by modulating the state of the local network, they can shape responses to incoming inputs as observed \textit{in vivo}. Such waves also occur in motor areas and~\citet{linden_movement_2022} have recently presented ensemble-recordings of neurons in the lumbar spinal cord that indicate that, rather than alternating, the population is performing a low-dimensional ``rotation'' in neural space, in which the neural activity is cycling through all phases continuously during the rhythmic behavior.

Interestingly, it can be shown that these travelling waves could have a measurable impact on the activity of the visual cortex. This has been illustrated in a recent study investigating the long-range apparent motion effect (lrAM)~\citep{chemla_suppressive_2019}. The lrAM is the simple phenomenon of perceiving a smooth motion when showing two dots in a temporal sequence and in relatively close visual proximity. The lrAM is the core building block underlying the use of sequences of images to induce the perception of smooth, realistic visual scenes, which is at the base of movies seen in cinema theaters. In this study, the authors used voltage-sensitive dye imaging to record the activity of the primary visual cortex of macaque monkeys to the presentation of the pair of dots presented independently or in conjunction. A probabilistic modelling has shown that the activity of the joint presentation induced a suppressive wave in the direction opposed to the perceived direction, shaping the formation of a wave of propagation travelling at a speed compatible with the perceived motion. A computational model validated the hypothesis that this process could be mediated by diffusion in the horizontal layers connecting the different locations within this cortical area. In summary, the study by~\citet{chemla_suppressive_2019} gave a multi-disciplinary account to demonstrate the effect of travelling waves in the visual cortex.
%
% Robust computation with rhythmic spike patterns.~Proceedings of the National Academy of Sciences of the United States of America~116(36), 18050 - 18059.~ https://dx.doi.org/10.1073/pnas.1902653116
%
%%%%%%%%%%%%%%%%%%%%%%%%%%%%%%%%%%%%%%%%%%%%%
\subsection{A rediscovered hypothesis: precise spiking motifs in cell assemblies}
%%%%%%%%%%%%%%%%%%%%%%%%%%%%%%%%%%%%%%%%%%%%%
%%%%%%%%%%%%%%%%%%%%%%%%%%%%%%%%%%%%%%%%%%%%%
Travelling waves indicate that spatio-temporal correlations could play an important role in shaping neural activity. For instance, statistical dependencies in the responses of sensory neurons govern both the amount of stimulus information conveyed and the means by which downstream neurons can extract it. In particular, this was put in evidence by analyzing the functional significance of correlated firing in a complete population of macaque parasol retinal ganglion cells using a model of multi-neuron spike responses~\citep{pillow_spatio-temporal_2008}, which shows precise spatio-temporal differences in this recurrently connected assembly. The different aspects of information in the data are evaluated by a decoding strategy, highlighting the role of correlations. Note that a similar dataset used in~\citep{schneidman_weak_2006} is available from Michael Berry's lab~\citep{berry_spike_2022} and allows \change[Antoine]{performing and validate}{testing in order to validate or falsify} these hypotheses. However, \remove[Antoine]{a limit of explaining such data with travelling waves is that} in theory, a \add[Antoine]{cortical} travelling wave would be stationary, which is incompatible with the limits in space and time of a neural system. Recent observations may suggest that neural groups or ensembles, rather than individual neurons, are emergent functional units of cortical activity. ~\citet{miller_visual_2014} show that whereas intrinsic ensembles recur at random time intervals, visually evoked ensembles are time-locked to stimuli. Experiments are performed using two-photon calcium imaging of populations of neurons from the primary visual cortex of awake mice during visual stimulation and spontaneous activity. The study proposes that visual stimuli recruit endogenously generated ensembles to represent visual attributes. Note that evoked ensembles in response to a natural movie played in a loop were precisely timed across repetitions. % (Fig. 7).

From another viewpoint, there is a substantial literature in neurobiology indicating that brain dynamics often organize into stereotyped sequences such as synfire chains~\citep{ikegaya_synfire_2004}, packets~\citep{luczak_sequential_2007} or hippocampal sequences~\citep{pastalkova_internally_2008}. %
%%%%%%%%%%%%%%%%%%%%%%%%%%%%%%%%%%%%%%%%%%%%%
%\subsection{Explictly decoding spiking motifs}
%%%%%%%%%%%%%%%%%%%%%%%%%%%%%%%%%%%%%%%%%%%%%
%%%%%%%%%%%%%%%%%%%%%%%%%%%%%%%%%%%%%%%%%%%%%
%We have previously seen that neurobiologists have described stereotyped sequences of neural activation, notably in the hippocampus~\citep{pastalkova_internally_2008,villette_internally_2015,malvache_awake_2016}, and that neurons can be activated in the same order across days~\citep{haimerl_internal_2019}. 
Going further, researchers have found precise repetitions of spontaneous patterns of synaptic inputs in neocortical neurons, \textit{in vivo} and \textit{in vitro}. These patterns repeat after minutes, maintaining millisecond accuracy. Indeed,~\citet{ikegaya_synfire_2004} demonstrated that in cortical activity, one can find a repetition of several motifs in spike activity (duration around 1s +/- .5 s, some events in motifs are of similar size but sometimes absent). These sequences can be specific of a particular layer or column, can be synchronized with network activity oscillation, and can involve several cells. They also demonstrated that these sequences can form super sequences, so-called \emph{cortical songs}. It consists of the assembly of several sequences which repeat in a specific order with a compressed timing. \emph{In vivo} spontaneous activity also reveals repeating sequences: about 3000 sequences, each involving 3-10 cells out (of about 900), and lasting up to 3 seconds. Sequences have specific topographic structures, in some cases involving only a particular layer or a vertical column of cells or cells located in a cluster, and are associated with a structured spatial organization of the neurons that formed them. %In cortical songs, there is a ``compressing timing'' which may be taken into account by a similar mechanism as maxpooling in CNNs for space, but in time. Or there may be a mechanism for controlling the replay speed (pulvinar, \ldots{} , ?)
% %----------------------------%
% \begin{figure}
% \centering
% \includegraphics[width=.6\linewidth]{figures/Ikegaya2004zse0150424620001.jpeg}%pcbi.1006283.g001.PNG_L.png}
% \caption{Fig. 1. from~\citep{ikegaya_synfire_2004} repeated motifs of spontaneous synaptic activity \textit{in vitro} and \textit{in vivo}. (A) Repeated motifs of intracellular activity from layer 5 pyramidal neurons in slices. Panels show segments (red) of the same voltage-clamp recording from the same cell, repeating seconds or minutes after the initial occurrence (blue). Arrows indicate timings of repeated PSCs. (i) Upper trace: low--temporal resolution display of spontaneous activity of a neuron. Lower traces: higher resolution display of the repeated motif at indicated regions of the trace (a to c). (ii) Example of a longer motif. (B) Three repetitions of a motif. The top traces show the motifs superimposed on each other (blue, green, and red), the middle traces show these same traces individually, and the bottom traces show temporally magnified regions of the motifs (a to c). (C) Repeated sequences of intracellular current-clamp recordings \textit{in vivo}. Two (i) and three (ii) repetitions of motifs are shown. Shuffle tests were performed on traces (i), a to c, yielding significantly fewer repeats (fig.~S2, P  0.02). In (i), the blue trace is shifted --2.75 mV; in (ii), the blue trace is shifted --1.58 mV, and the green +0.79 mV.}\label{fig:Ikegaya2004}
% \end{figure}
% %----------------------------%

\begin{figure}
\centering
\includegraphics[width=\linewidth]{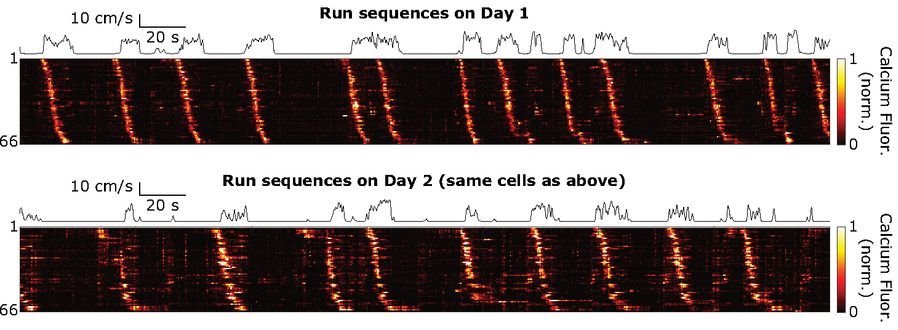}
\caption{\textbf{An example of a precise temporal motifs seen in cortical slices.} In this study by~\citep{haimerl_internal_2019}, an analysis of the raster plot shows repetitions of precise spiking motifs with a time scale of the order of seconds. (A) Calcium fluorescence (heatmap) of CA1 neurons participating to run sequences in consecutive imaging sessions. Cells have been selected and ordered with respect to their activity in the first imaging session (Top). The black line represents the speed of the mouse (modified from Figure 1 from~\citep{haimerl_internal_2019} under the CC-BY \href{https://www.pnas.org/doi/full/10.1073/pnas.1718518116}{PNAS License}).}\label{fig:haimerl}
\end{figure}

Additional studies detail the role of such precise spike timing in downstream information transfer and coding~\citep{villette_internally_2015,branco_dendritic_2010,luczak_packet-based_2015}. This is for instance relevant in sensory pathways in vision~\citep{meister_concerted_1995}, audition~\citep{decharms_primary_1996}, olfaction~\citep{cleland_construction_2014, kashiwadani_synchronized_1999, rinberg_speed-accuracy_2006} or touch~\citep{johansson_first_2004}.  In particular, stereotyped sequences of neural activation have been described in the adult hippocampus and related to its function in mental travel in time and space~\citep{buzsaki_space_2018}. These sequences can be internally generated~\citep{pastalkova_internally_2008,villette_internally_2015} and may be formed by the chained activation of orthogonal assemblies, themselves organized as sequence packets~\citep{malvache_awake_2016}. In that protocol, hippocampal sequences are formed by the ordered activation of smaller sequence motifs. They are stereotyped and robust since neurons can be activated in the same order across days (see Figure~\ref{fig:haimerl} from~\citep{haimerl_internal_2019}). As a consequence, hippocampal sequences may rely on an internally hardwired structure and form the functional building blocks for encoding, storing and retrieving experience.
%
%----------------------------%
%----------------------------%

It is interesting to make a parallel with the ``Rapid Formation of Robust Auditory Memories'' reported in~\citep{agus_rapid_2010} which uses noise patterns to observe if listeners could learn to detect repeated occurrences of some frozen noise patterns. In particular, they used random waveforms to probe the formation of new memories for arbitrary complex sounds. A behavioral measure was designed, based on the detection of repetitions embedded in noises up to $4 \si{\second}$ long. %The task is to detect the repetition of the same (frozen) noise within a trial.
Unbeknownst to listeners, some noise samples reoccurred randomly throughout an experimental block. They showed that the ``repeated exposure induced learning for otherwise totally unpredictable and meaningless sounds'' by showing that the sensitivity increases in that case. Note also that ``acoustical analyses failed to reveal any obvious differences between good and bad noises'' and that ``time reversal had no significant effect on the detection accuracy (which is quite surprising). The learning is unsupervised (statistical, automatic), fast-acting (phase transition, ``insight''), and long-lasting (memorization). 

That results suggest that precise spiking motifs are not necessarily grouped on a topography and can be forming apparently randomly arranged connections. Interestingly, one theoretical viewpoint considers synfire braids~\citep{bienenstock_model_1995}, where a precise sequential motif of spikes will synchronize as it reaches the soma of a neuron for which synaptic delays are adequately tuned. Furthermore, computational modeling shows that at the scale of neurons, an efficient neural code can emerge where spike times are organized in prototypical, precise temporal motifs~\citep{izhikevich_polychronization_2006} which this author defined as \emph{polychronous groups}. The rest of this review will be devoted to present evidence for the use of such precise spiking motifs in computational neuroscience, neurobiology and neuromorphic engineering. As a summary, it seems that such precise structural information is essential to the neural code and that it seems imperative to include this information in decoding algorithm for a better understanding of neural activity.
%
% TODO: re-read~\citep{luczak_packet-based_2015} Luczak A, McNaughton BL, Harris KD. Packet-based communication in the cortex. Nat Rev Neurosci. 2015;16(12):745--55.
% \subsubsection{Neural Variability and Sampling-Based Probabilistic Representations in the Visual Cortex~\citep{orban_neural_2016}}
% \begin{itemize}
%  \item
%   Stochastic sampling links perceptual uncertainty to neural response variability
% \item
%   Model accounts for independent changes in strength and variability of responses
% \item
%   Model predicts relationship between noise, signal, and spontaneous correlations
% \item
%   Stimulus statistics dependence of response statistics is explained
% \end{itemize}
%%%%%%%%%%%%%%%%%%%%%%%%%%%%%%%%%%%%%%%%%%%%%
%%%%%%%%%%%%%%%%%%%%%%%%%%%%%%%%%%%%%%%%%%%%%
\section{Understanding precise spiking motifs in neurobiology}\label{sec:detection}
%%%%%%%%%%%%%%%%%%%%%%%%%%%%%%%%%%%%%%%%%%%%%
%%%%%%%%%%%%%%%%%%%%%%%%%%%%%%%%%%%%%%%%%%%%%
%%%%%%%%%%%%%%%%%%%%%%%%%%%%%%%%%%%%%%%%%%%%%
\subsection{Decoding neural activity from firing rates}
%%%%%%%%%%%%%%%%%%%%%%%%%%%%%%%%%%%%%%%%%%%%%
In this section, we will review current evidence on how we may take advantage of spiking motifs in neurobiology, that is, in an effort to understand actual recordings from biological neural tissues. In most generic computational models, the neural activity is assumed to be encoded in the firing rate. For instance, the output of the so-called linear non-linear (L-NL) models is assumed to model the response of a biological neuron as the sequence of a linear integration followed by a non-linear spiking response generating spikes according to a Poisson point-process\note[Laurent]{adding reference}~\citep{simoncelli_characterization_2003}. As such, a simple decoding strategy is to infer the input knowing the neuron's tuning curves, that is, its selectivity to a range of features~\citep{jazayeri_optimal_2006} or simply by a simple regression~\citep{berens_fast_2012}. This latter model assumes a Bernoulli model for the generation of spikes, such that the decoding amounts to a single-layer logistic regression. An important perspective of these methods used to decipher the recorded activity is that it could be ultimately used to fit neural network models to the recorded activity~\citep{bellec_fitting_2021}. In this particular paper, authors fit summary statistics of neural data with a (differentiable) spiking network model. The loss function is the cross entropy (following Bernouilli hypothesis with a GLM where each unit is modelled with an SRM neuron~\citep{gerstner_time_1995}) and embedded with recurrent dynamics. In particular, it comes with \href{https://github.com/EPFL-LCN/pub-bellec-wang-2021-sample-and-measure}{code} and uses the \note[Laurent]{now citing the URL}publicly available V1 dataset~\citep{kohn_utah_2016} which allows supervising the model with the input being the movie and the output the spikes recorded. These type of model may infer sparse activity in a set of binary latent variables, each describing the activity of a cell assembly~\citep{warner_probabilistic_2022}.
\change[Antoine]{Using the right metric, as implemented in that paper by the corresponding neural models, is essential in better understanding neural data.}{Carefully picking the more appropriate metric, as implemented in that paper by the corresponding neural models, is essential in better understanding neural data.} Importantly, these models are dependent on a core definition of spike measures, and we will review here how precise spiking motifs are taken into account by such spike distances.
%
 %%%%%%%%%%%%%%%%%%%%%%%%%%%%%%%%%%%%%%%%%%%%%
\subsection{Decoding neural activity using spike distances}
 %%%%%%%%%%%%%%%%%%%%%%%%%%%%%%%%%%%%%%%%%%%%%
There are different solutions to provide with a distance between two given spike trains. A known measure is the Victor-Purpura distance, which overcomes inconsistencies experienced with a \remove[JN]{simple} firing rate (Poisson model) of spike trains~\citep{victor_nature_1996}. \change[JN]{In a later study, Van Rossum (2001) have defined a distance that uses a time constant as a parameter}{Then a study tries to solve the problem by including a time constant as a parameter}~\citep{van_rossum_novel_2001}. \change[JN]{Depending on this parameter, the distance interpolates between a coincidence detector and a rate difference counter.}{This parameter will then be used to interpolate the distance between a coincidence detector and a rate difference counter.} Such distances were extended to non-Euclidean metrics and use morphological manipulations to compute spike train dissimilarity~\citep{kreuz_measuring_2007}. Mathematically, the stability of distance measures induced by level-crossing sampling can be evaluated~\citep{moser_stability_2014}, notably in light of the so-called Weyl's discrepancy measure~\citep{weyl_ueber_1916}, which may lead to the definition of a cross-correlation measure \change[JN]{which}{. An interesting conclusion since cross-correlation measure is} which is adapted to the event-based nature of spiking signals. These observations lead to the intuition that each distance may be as well  as the optimal solution of a generative model for these measures, possibly through non-linear relations~\citep{aronov_non-euclidean_2004}. % ~\citep{} 

\begin{figure}
\centering
\includegraphics[width=\linewidth]{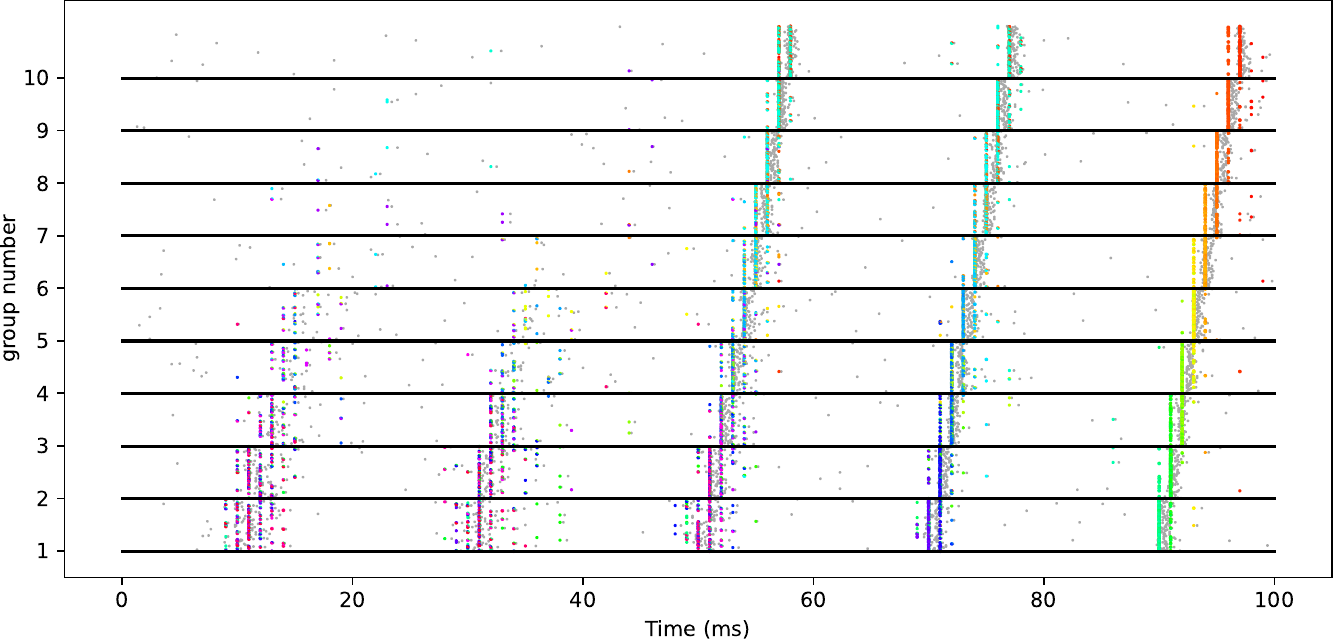}
\caption{\textbf{Detecting motifs using SPADE.} We used the SPADE algorithm~\citep{stella_3d-spade_2019} by adapting their \href{https://elephant.readthedocs.io/en/latest/tutorials/spade.html}{tutorial} on the data generated in  Figure~\ref{fig:diesman}. This allowed to label different precise spike motifs which are denoted by different colors. Spikes belonging to the same motif have the same color. %As the synfire chain gets more precise, the detected patterns grow in size.
}\label{fig:Diesmann_et_al_1999_spade}
\end{figure}

%Let us now focus on distances defined based on a generative model formed by using precise spiking motifs. 
Concerning spike timings,~\citet{levakova_review_2015} \change[Antoine]{has}{have} \change[JN]{reviewed the methods for neuronal response latency estimation exists and includes Bayesian binning}{reviewed existing methods for estimating the latency of neural responses that include Bayesian binning.}. \change[JN]{More broadly }{Alternatively }, unitary event analysis can be performed by a statistical model of coincidence detection~\citep{grun_unitary_2002-1}. This was extensively used in detecting above chance significant synchronous patterns~\citep{grun_unitary_2010}, in particular in recordings of pairs of neurons (see~\citep{riehle_spike_1997} for instance), and has been extended to non-stationary data~\citep{grun_unitary_2002}. A method to detect significant patterns of synchronous spiking in a subset of massively parallel spike trains in the presence of background activity can be defined using the statistical evaluation of synchronous spike patterns extracted by frequent item set mining~\citep{torre_statistical_2013}. By the same group, the SPADE, CAD or ASSET algorithms are methods for identification of spike patterns in massively parallel spike trains (the spiking activity of tens to hundred(s) of neurons recorded in parallel) by identifying fine temporal correlations in the$~\ms$ precision range~\citep{quaglio_methods_2018}. This was recently extended in~\citep{stella_3d-spade_2019} in order to find re-occurring patterns in parallel spike train data, and to determine their statistical significance. The extension improves the performance in the presence of patterns with different durations, as demonstrated by application to various synthetic data \add[Laurent]{such as the synthetic data for synfire chains (see Figure}~\ref{fig:Diesmann_et_al_1999_spade}), such as surrogates generated to evaluate precisely timed higher-order spike correlations~\citep{stella_comparing_2022}.

Another important algorithm, called SPOTDisClust, is based on the detection of structured temporal patterns~\citep{grossberger_unsupervised_2018}. They introduced an unsupervised method based for their detection from high-dimensional neural ensembles. The algorithm measures similarity between two ensemble spike patterns by determining the minimum transport cost of transforming their corresponding normalized cross-correlation matrices into each other. Many approaches to this problem are supervised. In other words, they take patterns occurring concurrently with a known event, such as the delivery of a stimulus for sensory neurons or the traversal of a running track for determining hippocampal place fields, as a ``template'' and then search for repetitions of the same template in spiking activity~\citep{nadasdy_replay_1999,lee_combinatorial_2004}. In SPOTDisClust, the learning is unsupervised. It uses the prior that there is only one spike per pattern. Using a so-called t-SNE projection (that allows to project this high-dimensional representation to a lower-dimension map) validated that this clustering method can retrieve all patterns from the data. % (see Figure~\ref{fig:SPOTDisClust}). 
The limits of this method are that it is computationally complex, block-based and strictly specialized for the task at hand. To overcome these difficulties, a novel method was recently developed~\citep{sotomayor-gomez_spikeship_2021}.
%   Davidson TJ, Kloosterman F, Wilson MA. Hippocampal replay of extended experience. Neuron. 2009;63(4):497--507. pmid:19709631
\add[JN]{Whether it is the distance between two given spike trains or a comparison of the spike timings, the complexity and the diversity of the methods used to measure them are witnesses of the growing interest of the integration of these measures in the understanding of the neural code. One of the steps to test their potential usefulness is then to scale these methods to larger amounts of data.} 

%----------------------------%
%----------------------------%

% %%----------------------------%
% \remove[LP]{
% \begin{figure}
% \centering
% \includegraphics[width=\linewidth]{figures/pcbi.1006283.g001.png}%pcbi.1006283.g001.PNG_L.png}
% \caption{\textbf{The SPOTDisClust algorithm.} Fig 1 of~\citep{grossberger_unsupervised_2018}, showing simulated example illustrating the steps in the algorihtm. \textbf{(A)} Structure of five ``ground-truth'' patterns (\ldots). For each pattern and each neuron, a random position was chosen for the activation pulse. \textbf{(B)} Neuronal output is generated according to an inhomogeneous Poisson process, with rates dictated by the patterns in (A). (© Authors under a \href{https://journals.plos.org/ploscompbiol/article?id=10.1371/journal.pcbi.1006283}{CC licence})}\label{fig:SPOTDisClust}
% \end{figure}
% %----------------------------%
% }
%
%%%%%%%%%%%%%%%%%%%%%%%%%%%%%%%%%%%%%%%%%%%%%
\subsection{Scaling up to very large scale data}
%%%%%%%%%%%%%%%%%%%%%%%%%%%%%%%%%%%%%%%%%%%%%
Over the past decade, tremendous technological advances across several disciplines have dramatically expanded the frontiers of experimentally accessible neuroscientific facts. Bridging across different spatial and temporal scales, combination of \textit{in vivo} two photon imaging, large population recording-array technologies, optogenetic circuit control tools, transgenic manipulations as well as large volume circuit reconstructions are now used to examine the function, structure and dynamics of neural networks on an unprecedented level of detail and precision. The daunting complexity of the biological reality revealed by these technologies highlights the importance of neurobiological knowledge to provide a conceptual bridge between abstract principles of brain function and their biological implementations within neural circuits. As a consequence, there is a growing need to scale these methods to larger amounts of data. 

There are multiple \change[Antoine]{solutions}{approaches} which aim at tackling this problem. One algorithm capable of achieving such a daunting task is the Rastermap algorithm~\citep{pachitariu_robustness_2018}. Basically, it rearranges neurons in the raster plot based on the similarity of their activity and applies a deconvolution strategy based on a linear model. \add[Laurent]{Yet this method was mainly tested on calcium imaging data which is known to add some imprecision to the timing of the original spiking activity.} The model is \note[Laurent]{now citing the URL}openly accessible~\citep{stringer_mouselandrastermap_2020} and has led to important discoveries. In~\citep{stringer_spontaneous_2019} for instance, it was shown that a neuronal population encodes information most efficiently when its stimulus responses are high-dimensional and uncorrelated, and most robustly when they are lower-dimensional and correlated. Then, in~\citep{stringer_high-precision_2021}, the authors analyzed spontaneous neural firing, finding that neurons in the primary visual cortex encoded both visual information and motor activity related to facial movements. In~\citep{russo_cell_2017}, the authors developed novel machine learning tools and statistical tests for unsupervised spatio-temporal pattern detection in non-stationary environments, which were applied to simultaneous electro-physiological recordings from tens to hundreds of neurons for decoding cognitive processes from neural activity. Altogether, this provides evidence for the importance of such machine-learning-based tools to provide with breakthroughs in neuroscience.
%   Data availability: All of the processed deconvolved calcium traces are available on \href{https://figshare.com/articles/Recordings_of_ten_thousand_neurons_in_visual_cortex_in_response_to_2_800_natural_images/6845348}{figshare}, together with the image stimuli.

% \subsection{Extending the decoding at multiple time scales}
In the paper by~\citet{russo_cell_2017}, the authors present a unifying methodological and conceptual framework which detects assembly structure at many different time scales, levels of precision, and with arbitrary internal organization. It uses  sliding window as in~\citep{grun_unitary_2002} and the reliable and efficient analysis of an excess or deficiency of joint-spike events~\citep{pipa_neuroxidence_2008}. They extend the measure to multiple lags~\citep{torre_synchronous_2016}. The core measure is based on a non-stationarity corrected parametric statistical test for assessing the independence of pairs and an agglomerative, heuristic clustering algorithm for fusing significant pairs into higher-order assemblies.  To overcome the limits of models which either require spike times to be discretized, utilize a suboptimal least-squares criterion, or do not provide uncertainty estimates for model predictions or estimated parameters,~\citep{williams_point_2020} address each of these shortcomings by developing a point process model that characterizes fine-scale sequences at the level of individual spikes and represents sequence occurrences as a few marked events in continuous time. \change[Laurent]{They also introduce}{As originally introduced by}~\citep{kass_statistical_2005}, they use learnable time-warping parameters to model sequences of varying duration, which have been experimentally observed in neural circuits, and demonstrate these advantages on experimental recordings from songbird higher vocal center and rodent hippocampus.
At a larger scale, in~\citep{van_kempen_top-down_2021}, it was shown that attentional information from V4 or arousal can change the timings of groups of events in V1. They develop a Hidden Markov Model for quantifying the transitions. In particular, they show that fluctuations in neural excitability are coordinated between visual areas with retinotopic precision. Top-down attention drives inter-areal coordination along the reverse cortical hierarchy, predicting better behavioral performance with increased coordination. Building such models for predicting changes of timings based on context, such as using a so-called change point model for blocked-based experimental protocols~\citep{pasturel_humans_2020}, could therefore provide useful prior information to enhance the decoding from neural activity.
% sparse in time and space : AL Barth and JF Poulet Trends in Neurosciences 35.6 (2012), pp.~345-355.  CC Petersen and S Crochet, Neuron 78.1 (2013), pp.~28-48.
%
%%%%%%%%%%%%%%%%%%%%%%%%%%%%%%%%%%%%%%%%%%%%%
%%%%%%%%%%%%%%%%%%%%%%%%%%%%%%%%%%%%%%%%%%%%%
\section{What biological mechanism could allow learning spiking motifs?}
%%%%%%%%%%%%%%%%%%%%%%%%%%%%%%%%%%%%%%%%%%%%%
%%%%%%%%%%%%%%%%%%%%%%%%%%%%%%%%%%%%%%%%%%%%%
%
Despite the evidence for the effectiveness of precise spiking detection we presented above, doubts may remain as to the reliability of this learning mechanism and whether there is a real need for further research on this subject. The discovery of the existence of an equivalent biological mechanism in vertebrates as well as the demonstration of the importance of its role in various developed behaviors allow us to put these doubts to rest. In the following paragraphs we will successively describe the first biological observations of delay learning, identify myelinization as an important actor and finally study this phenomenon at the molecular level. 

%%%%%%%%%%%%%%%%%%%%%%%%%%%%%%%%%%%%%%%%%%%%%
\subsection{Biological observations of delay adaptation}
%%%%%%%%%%%%%%%%%%%%%%%%%%%%%%%%%%%%%%%%%%%%%
% First biological observations

One of the first significant evidence of any neuronal delay in the information propagation within the animal neural system came from Hermann von Helmholtz's study of a frog's sciatic nerve in 1850~\citep{von_helmholz_messungen_1850,peyrard_how_2020}, and was later confirmed with Young's study of the squid giant axon~\citep{young_functioning_1938}. Dendritic propagation delays vary from sub-milliseconds to a few milliseconds, while axonal propagation delays range from a few milliseconds to tens of milliseconds, depending on the neuronal population studied~\citep{madadi_asl_dendritic_2018}. Extensive measures of nerve conduction velocities were done in different animals, including humans, and significant variations related to age, sex and other factors have been identified~\citep{stetson_effects_1992}.

However, it was not until the study of the interaural time difference (ITD) by \remove[Amélie]{gerstnerneuronal1996}~\citet{carr_circuit_1990} that it was discovered that this delay is not homogeneous for all neurons of the same type and species, but adapts according to their function. This ITD is a biological mechanism which allows for the azimuthal localization of sound by barn owls, by organically computing the difference in arrival time of a sound between their two ears. \add[Amélie]{It was first theorized in the Jeffress "ITD-versus-place model"}~\citep{jeffress_place_1948}. \add[Amélie]{As hypothesized by this model, the \textit{nucleus laminaris} of the avian brain contains coincidence detectors and, associated with the \textit{nucleus magnocellularis} axons, forms circuits for processing ITD}~\citep{konishi_coding_2003}. According to~\citep{gerstner_neuronal_1996}, there is a true paradox in auditory neural systems since ``neural networks encode behaviorally relevant signals in the range of a few $\mu s$ with neurons that are at least one order of magnitude slower'', therefore necessarily need to play on the response time to do so. \add[Amélie]{This assertion confused the mean interspike interval, i.e. how often a neuron can fire, and the specific spike time, i.e. how precisely a spike can be generated. However, it has nevertheless contributed to the recognition of the importance of time in various biological mechanisms.} \remove[Amélie]{penacochlear2001 conducted an extended study on ITD in order to confirm this theory by identifying the coincidence detector involved, which the authors situated in the nuclear laminaris. In the mouse somatosensory cortex, camontiming2019 found a strong correlation between the delay of the mouse behavioral response and the timing of multiunit activity evoked by a trained whisker.} A first hypothesis suggested that the sound coincidence was detected using stereausis, i.e. the temporal disparity between left and right cochlear loci in the owl's brain. However, it was quickly set aside as the predictions did not match the measured disparities in the loci, and no variation was perceived in the nuclear laminaris for a similar sound intensity in both owl ears. The authors support a second hypothesis, that of different axonal delays in the ipsi- and contra-lateral cochlear nucleus magnocellularis.~\citet{seidl_mechanisms_2010} experimentally seconds this hypothesis of a ``coarse'' regulation of delay, as the authors conclude that regulations at different sites within individual axons of at least two parameters, namely the axon diameter and internode distances, might be responsible for precise adjustments of physiological delays thus allowing the ITD detection. \add[Amélie]{The authors also note that the barn owl's axons seem to change in length, thus implementing a ``pure delay line''.}

\add[Amélie]{The experiments described above thus conclude on the important role of physiological delays in the avian sound localisation behaviour. However, the relevance of precise timing in spikes is not limited to birds; for example, in the mouse somatosensory cortex,}~\citep{camon_timing_2019} \add[Amélie]{found a strong correlation between the delay of the mouse behavioral response and the timing of multiunit activity evoked by a trained whisker.} These experiments also confirmed previous studies stating that the conduction velocity of a spike in a neuron (in other words, its delay) depends strongly on the axon diameter~\citep{gasser_axon_1939} and the internode distance between Ranvier nodes~\citep{brill_conduction_1977}\change[Amélie]{--- instead of solely the axon's length as was previously thought.}{. This mechanism adds to the axonal length delay, which was previously thought sole influence on the conduction velocity due to its anatomical soundness --- as}~\citet{seidl_mechanisms_2010}\add[Amélie]{ experimentally demonstrated, this mechanism by itself is not sufficient to explain the biological functionality but should be added to the one of axonal length delay.}

%%%%%%%%%%%%%%%%%%%%%%%%%%%%%%%%%%%%%%%%%%%%%
\subsection{The importance of myelination}
%%%%%%%%%%%%%%%%%%%%%%%%%%%%%%%%%%%%%%%%%%%%%
% The importance of myelination
\citet{gasser_axon_1939} experimentally confirmed with homogeneously selected neurons that the axonal delay is positively proportional to \change[Amélie]{both the axon diameter by itself and the diameter of the axon wrapped in myelin}{the axon diameter, i.e. the amount of myelin wrapped around the axon}. Indeed, the oligodendrocytes, one of the many glial cells present in the Vertebrates' vertebrates' nervous system identified in 1924 by Pío del Rigo Hortega~\citep{perez-cerda_pio_2015}, produce thin protein sheets interspersed with lipid layers wrapped concentrically around the axon, called myelin~\citep{schmitt_ultrastructure_1939}. Myelinization consists in ``two motions: the wrapping of the leading edge of the inner tongue around the axon underneath the previously deposited membrane and the lateral extension of myelin membrane layers toward the nodal regions''~\citep{simons_oligodendrocytes_2016}. Multiple myelin regions can appear on one neuron axon and form the following subdomains: ``the internode (corresponding to the compacted region of myelin), the paranodes (where the outer loops of the myelin contact the axon), the juxtaparanode (the interface between the paranode and compact myelin, rich in potassium channels) and the node of Ranvier (the approx $1~\mu m$ gap between adjacent myelin internodes [allowing for] the saltatory conduction)''~\citep{duncan_neuron-oligodendrocyte_2021}. \change[Antoine]{In}{On} average, each oligodendrocyte produces 20 to 60 myelinization processes and each myelin sheath is 20 to 200 $\mu m$ long~\citep{simons_oligodendrocytes_2016}. This demonstrates an additional impact of myelinization on the conduction velocity, as the number of segments is positively correlated to the axonal delay~\citep{brill_conduction_1977}. Thus, in several occasions, myelin has been identified as \change[Antoine]{a main}{an important} actor in the regulation of conduction velocity in neurons, i.e. axonal delay regulation.

% role of myelination and neuronal delay
\citet{fields_new_2015, fields_myelin_2020} state that myelin facilitates both the neural circuit function and the behavioral performance: experiments on mammals show that myelinization is activity-dependent and directly related to learning and memory consolidation, especially sensory or motor training and in enriched environments. This biological phenomenon takes place both at an early age, where the amount of oligodendrocites is particularly high in the central nervous system~\citep{reynolds_study_1928}, and in older animals, due to its involvement in coupling the activity of distant neuron populations. Myelination helps memory consolidation by coupling the activity of distant neuron populations and generating nearly synchronous responses in postsynaptic neurons involved among others in path integration~\citep{madadi_asl_dendritic_2018}, as was experimentally demonstrated on mice using a Morris water maze~\citep{steadman_disruption_2020}, contextual fear conditioning~\citep{pan_preservation_2020} or oligodendrocyte precursor cells (OPCs) knock-out~\citep{wan_impaired_2020}. Myelin also inhibits axon sprouting and synapse formation, especially in pyramidal neurons~\citep{fields_new_2015}, and is involved in axonal energy saving through a reduced axonal capacitance and a shift of the metabolic load from axons onto oligodendrocytes~\citep{duncan_neuron-oligodendrocyte_2021}.

It is worth highlighting that myelinization becomes increasingly important in larger brains where conductance delays are substantial and brainwave rhythms are critical; synchrony errors can lead to neuropsychiatric and neurological dysfunctions~\citep{fields_new_2015}, such as Parkinson's disease, epilepsy or multiple sclerosis~\citep{madadi_asl_dendritic_2018}. Additionally, a recent study suggests that demyelination of the optic nerve could be an underlying factor in glaucoma~\citep{xue_demyelination_2021}. ~\citet{duncan_neuron-oligodendrocyte_2021} states that ``the loss of myelin and oligodendrocytes fundamentally alters the neuron, [which are then] susceptible to energetic failure [and] subsequent degeneration''.  

%%%%%%%%%%%%%%%%%%%%%%%%%%%%%%%%%%%%%%%%%%%%%
\subsection{Interplay of delay adaptation and neural activity}
%%%%%%%%%%%%%%%%%%%%%%%%%%%%%%%%%%%%%%%%%%%%%
% How is the neuronal activity detected ?
However, one question remains: how do oligodendrocytes detect neuronal activity and regulate the myelinization accordingly? To answer this, we must first study the myelinization process. The OPCs first proliferate in the white matter \textit{via} a self-repulsive process, thus allowing for an evenly spaced network, and identify target axons. Most OPCs then differentiate into oligodendrocytes and immediately initiate myelinization, with no further migration~\citep{simons_oligodendrocytes_2016}. Not much is known about how oligodendrocytes select the axons to myelinate, but it seems that myelination only takes place on large enough axons and is strongly regulated by several factors~\citep{kuhn_oligodendrocytes_2019}, such as $Ca^{2+}$ activity~\citep{baraban_ca2_2018} of the neuregulin 1 growth factor~\citep{nave_axonal_2006}. The important role of myelination on delay learning and biological behaviors, as highlighted in the previous paragraph, suggests that the identification of target neurons as well as the myelin production is also regulated by neuronal activity. Indeed,~\citet{cullen_periaxonal_2021} experimentally demonstrated that learning and associated neuronal activity modify the Ranvier nodes' length and the periaxonal space width in the adult mouse brain. They also confirmed that the delay correlates with the level of skill acquisition.~\citet{gibson_neuronal_2014} suggests that neuronal activity does not solely promote adaptive myelination in the mammalian brain, but also OPC differentiation and oligodendrogenesis. Some further studies show that oligodendrocytes may detect neuronal activity thanks to growth factors or neurotransmitters released through ion channels or \textit{via} exocytosis, but does not require any axo-glial synaptic communication~\citep{fields_new_2015}. 

A potential scenario for selective myelination on electrically active axons using non-synaptic junctions between an axon and an oligodendrocyte is: the axon releases glutamate in the extracellular environment by vesicle fusion, which activate the oligodendrocyte's NMDA and metabotropic glutamate receptors. This triggers the axo-glial signalling complex, involving the phosphorylation of the SRC family kinase FYN followed by the translation of heterogeneous nuclear ribonucleoprotein A2 into local myelin basic proteins~\citep{fields_new_2015}.

% myelination for axonal delay, but what about dendritic delay ? 
The previous paragraphs present the biological mechanisms behind the axonal delay, regulated by myelination. However, the dendritic delay \add[Amélie]{as well as the axonal delay (see }~\citep{seidl_mechanisms_2010}) seem to play an equally important role in the precise timing within a sequence of spikes. Dendritic delay is involved in the performance, structure and function of the nervous system, the modulation of spatio-temporal properties of pre- and post-synaptic activity patterns and the functional limitations of sensory feedback control efficiency~\citep{madadi_asl_dendritic_2018}. Its role has been specifically identified in the compensation of input asynchrony in the mammalian auditory brain stem~\citep{spencer_compensation_2018}. \citet{mel_synaptic_2017} \add[Amélie]{highlights the dendrites' impact on neuronal plasticity, which is caused by the wide variation of numerous parameters: plasticity rules applied to different dendritic subtrees or dendritic subregions, local passive cable properties, distance travelled by remote dendritic inputs, branching structures, dendritic diameters, the relative timing of back-propagating somatic action potentials, etc. Dendritic spiking involvement in synaptic potentiation following active backpropagation into dendrites was experimentally uncovered using calcium imaging to highlight dendritic calcium entry allowing for long-tem potentiation}~\citep{golding_dendritic_2002}. ~\citet{branco_dendritic_2010} demonstrates the dendritic sensitivity to a sequence of synaptic activation in cortical pyramid neurons, encoded by ``both local dendritic calcium signals and somatic depolarization, leading to sequence-selective spike output''. The dendritic mechanism described can identify patterns delivered to a single dendrite or randomly distributed across the dendritic tree and relies on the dendritic calcium influx moderation by NMDA receptors.

All in all, learning spike motifs requires significantly complex pathways and biological mechanisms, whether in the dendrites or the axon of the neuron. More and more is known about the non-trivial research topic that is biological delay learning, and extensive experimental data help develop ANNs, whose learning rules would be more neuromorphic. 
%
%%%%%%%%%%%%%%%%%%%%%%%%%%%%%%%%%%%%%%%%%%%%%
\section{Modeling precise spiking motifs in theoretical \& computational neuroscience}
%%%%%%%%%%%%%%%%%%%%%%%%%%%%%%%%%%%%%%%%%%%%%
%%%%%%%%%%%%%%%%%%%%%%%%%%%%%%%%%%%%%%%%%%%%%
%%%%%%%%%%%%%%%%%%%%%%%%%%%%%%%%%%%%%%%%%%%%%
%%%%%%%%%%%%%%%%%%%%%%%%%%%%%%%%%%%%%%%%%%%%%
Now that we have reviewed biological foundations for the role of delays in neural computations, we review, in the following section, theoretical models which directly take advantage of using precise spiking motifs. 
%%%%%%%%%%%%%%%%%%%%%%%%%%%%%%%%%%%%%%%%%%%%%
Spiking Neural Networks (SNNs)~\citep{maass_networks_1997} are natural candidates to use these precise temporal patterns in the brain. The approach which is currently most prominent in the SNN community is to use existing algorithms from machine learning and to adapt them to the specificity of spiking architectures~\citep{goltz_fast_2021} One such example is to adapt the successes of deep learning algorithms and to transfer the back-propagation algorithm to SNNs, as it is the most widely used to tune the weights of a classical (non spiking) Neural Network. In a nutshell, it considers the system as implementing an input / output function and iteratively updates the weights according to the direction and magnitude of the error's gradient. \change[Antoine]{Since the activation function of a spiking neuron is not differentiable,}{In deep learning, gradient is computed on the activation function and since spikes are not differentiable,} a recent popular approach consists in using a surrogate gradient~\citep{neftci_surrogate_2019} to ``cross-compile'' a classical Neural Network to a spiking architecture~\citep{rueckauer_conversion_2017}. SNNs reach in some case a similar performance as their non spiking equivalent, for instance on the \remove[Antoine]{N-}MNIST dataset for categorizing digits in a stream of events~\citep{susi_nmnsd-spiking_2021}. \change[Antoine]{The main reason for adopting this approach instead of classical architectures is the possibility to benefit from dedicated neuromorphic low-energy hardware. 
However, most biological neural systems are obviously more efficient than current state-of-the-art vision systems, both in terms of efficiency (accuracy), in speed (latency), and energy consumption.}{So far, this approach does not outperform classical architectures both in term of training efficiency and performances}~\citep{davies_loihi_2018}\add[Antoine]{. However, they remain the best candidates to reproduce biological neural systems and their capacities both in terms of accuracy, speed and energy consumption.} There is therefore an immense gap in the way we understand biology to translate it to the efficiency of SNNs. To go beyond the state-of-the-art, we will focus here on one core computation of a spiking neuron, that is, its ability to switch from the classical integrator mode (summing analog currents on its \change[Antoine]{synapses}{dendrites}) \change[Antoine]{to a synchrony detector where it emits a spike whenever presynaptic spiking inputs are synchronized}{to a detector of precise spiking motifs}~\citep{paugam-moisy_computing_2012}. In particular, we will explore different existing architectures which are able to overcome the diversity of input presynaptic patterns and learn to detect stable spiking motifs, that is, volleys of spikes which are stable up to a certain onset time (see Figure~\ref{fig:THC}). These models will be compared in light of neuroscientific and computational perspectives. %We review in this section theoretical and computational foundations of spiking motif detection. %In particular, our objective is to describe methods which fully exploit the capacity of spiking neurons to detect synchronous patterns.
%%%%%%%%%%%%%%%%%%%%%%%%%%%%%%%%%%%%%%%%%%%%%
\subsection{Izhikevich's polychronization model}
\begin{figure}%[t!]
  \centering
  \includegraphics[width=0.980\linewidth]{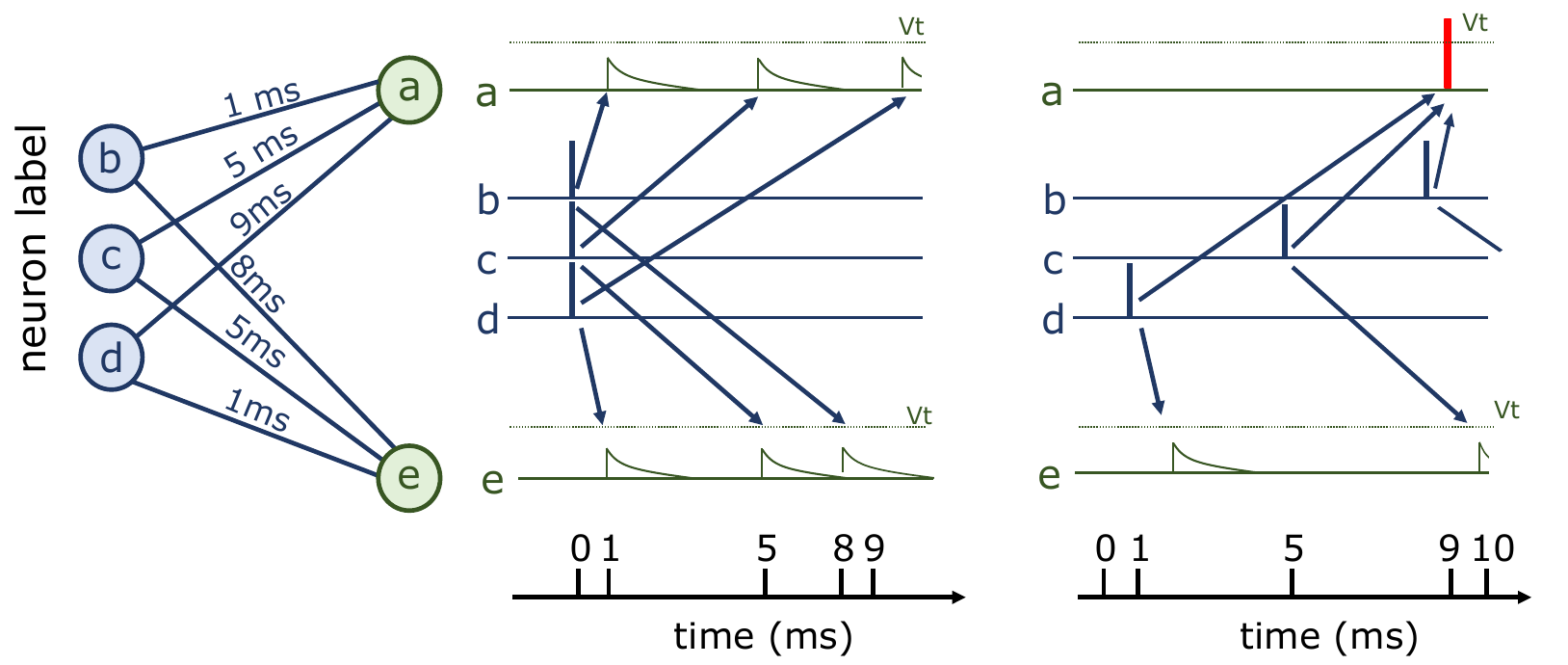}%png}% https://www.overleaf.com/5625872443qpcwrkssgbsf
    \caption{\textbf{Core mechanism of polychrony detection~\citep{izhikevich_polychronization_2006}.} \textit{(Left)}~In this example, three presynaptic neurons denoted \textit{b}, \textit{c} and, \textit{d} are fully connected to two post-synaptic neurons \textit{a} and \textit{e}, with different delays of respectively $1$, $5$, and $9~\ms$ for \textit{a} and  $8$, $5$, and $1~\ms$ for \textit{e}. \textit{(Middle)}~If three synchronous pulses are emitted from presynaptic neurons, this will generate post-synaptic potentials that will reach \textit{a} and \textit{e} asynchronously because of the heterogeneous delays, and they may not be sufficient to reach the membrane threshold in either of the post-synaptic neurons, therefore no spike will be emitted as this is not sufficient to reach the membrane threshold of the post synaptic neuron, so no output spike is emitted.
    %at these different delays, and these may not be sufficient to generate a spike in either neuron.
    \textit{(Right)}~If the pulses are emitted from presynaptic neurons such that, taking into account the delays, they reach the post-synaptic neuron \textit{a} at the same time (here, at $t=10~\ms$),  the post-synaptic potentials evoked by the three pre-synaptic neurons sum up, causing the voltage threshold to be crossed and thus to the emission of an output spike (red color), while none is emitted from post-synaptic neuron \textit{e}.
     }
  \label{fig:izhikevich}
\end{figure}

%%%%%%%%%%%%%%%%%%%%%%%%%%%%%%%%%%%%%%%%%%%%%
%%%%%%%%%%%%%%%%%%%%%%%%%%%%%%%%%%%%%%%%%%%%%
% polychronization
As we saw, most SNN, and in particular those adapted from analogous deep-learning-like architectures, rely on an encoding of information based on a continuously varying firing-rate. Notable exceptions of SNNs using precise spike timings are the \add[Antoine]{time encoding machine by}~\citet{lazar_time_2004} and \textit{polychronization} model of~\citet{izhikevich_polychronization_2006}. \add[Antoine]{In this section we focus on the polychronization model} based on a random recurrent model of spiking neurons including synaptic delays chosen from a range of biologically realistic delays (from $0$ to $20~\ms$) and whose weights evolved with a Spike-Time Dependent Plasticity (STDP) learning rule~\citep{markram_regulation_1997}. \change[Antoine]{It is consensus that spike timing (STDP) plays a crucial role in the development of synaptic efficacy for many different kinds of neurons}{It was shown that spike timing (STDP) has an impact on the development of synaptic efficacy for many kinds of neurons}~\citep{caporale_spike_2008}. Delays are defined as the total time taken for a spike to be conducted from one presynaptic neuron's soma to the efferent postsynaptic neuron's soma. It is worth mentioning that only the weights are changed using the STDP rule and that the set of delays is set randomly at initialization and that delays are then ``frozen'' for the rest of the simulation. Due to the interplay between the delays and STDP, the spiking neurons spontaneously self-organize into groups and generate patterns of stereotypical polychronous activity, i.e.~exhibit reproducible time-locked firing patterns which  the author defined as ``polychronous groups'' (PGs). One core ingredient of this model is the fact that the neurons composing a group fire at different times, but due to the heterogeneous delays, the spikes reach the postsynaptic neuron at the same time. This synchrony of arrival at the soma of the neuron leads to the summation of the excitatory post-synaptic potentials evoked by each spike, and thus to the crossing of the voltage threshold and to the discharge of a spike (see Figure~\ref{fig:izhikevich}). According to the STDP rule, the group of neurons involved in this polychronous activity will see their synaptic weight increase and, thus, may consolidate the formation of a polychronous group. 
%----------------------------%
%
% 
%----------------------------%

% follow-ups
Interestingly, the paper by~\citep{izhikevich_polychronization_2006} stirred a lively debate in the field of computational neuroscience, with a general positive acceptance, but relatively few works extended this seminal paper. Indeed, there were already existing models of synaptic delay learning in spiking neural networks, see for instance~\citep{huning_synaptic_1998} or~\citep{eurich_dynamics_1999} yet they had not shown potential applications to the detection of spiking motifs. A popular model for the detection of latency patterns is the \emph{tempotron}~\citep{gutig_tempotron_2006}, particularly reviewed in~\citep{gutig_spike_2014}. The \emph{tempotron} is a supervised synaptic learning algorithm which classifies a distractor from a target motif, in order to extend the perceptron, which does not incorporate a spike timing framework. The \emph{tempotron} learning rule is derived by an optimization process and takes the form of a supervised STDP rule. The limits of this model are that its output is only binary and that its storage capacities are limited. An extension of~\citep{izhikevich_polychronization_2006} was made in a very detailed work aiming at reproducing the polychronization model~\citep{pauli_reproducing_2018}. Indeed, while the original paper contained material within the text to reproduce the whole model (using MATLAB), it was not complete such as to allow for the reproduction of all results presented in that manuscript. This more recent work details how this code could be slightly corrected. It comes with a Python code and a version control system detailing the whole process used to give provenance to the different steps in this scientific process. Another recent work gives a Bayesian account in a similar model~\citep{guise_bayesian_2014}. In that work, based on the fact that previous methods for studying polychronous groups' activation response to stimuli have been limited by the template-based methods used to identify PG activation, the authors outline a new method that overcomes these difficulties by establishing a probabilistic interpretation of PG activation. They demonstrate the use of this method by investigating the claim that PGs might provide the foundation of a representational system. Stimulation of the trained network produces the activation of a PG, i.e. the propagation of firing activity through multiple layers due to convergent patterns of firing. While extending the original method, these methods reveal shortcomings that we will try to analyze in the rest of this section.

\begin{figure}%[t!]
  \centering
  \includegraphics[width=0.490\linewidth]{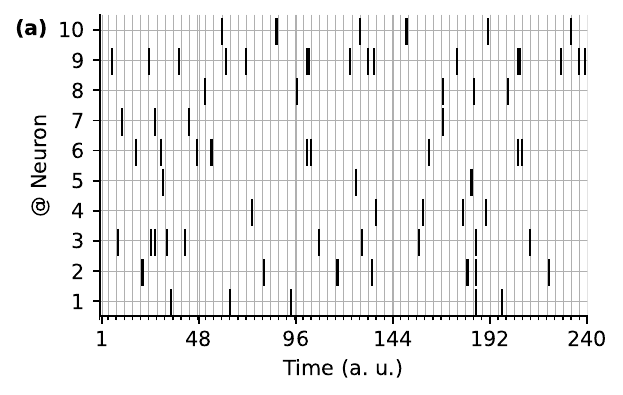}
  \includegraphics[width=0.490\linewidth]{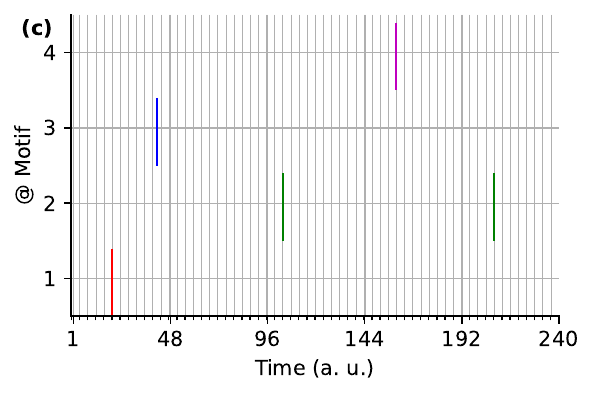}
  \includegraphics[width=0.490\linewidth]{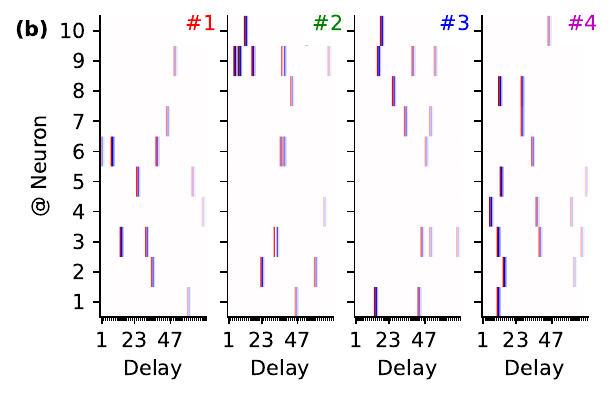}
  \includegraphics[width=0.490\linewidth]{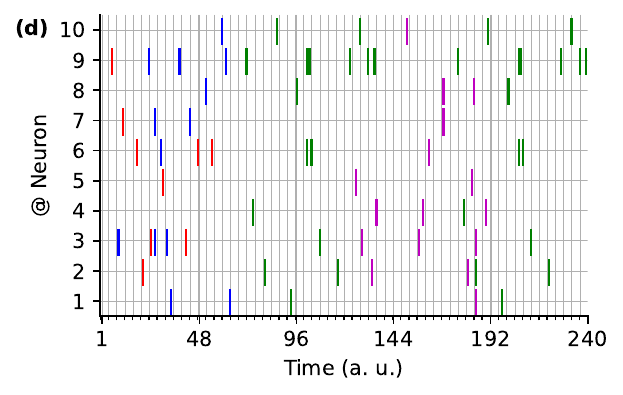}
    \caption{\textbf{Detecting event-based motifs using spiking neurons with heterogeneous delays.} 
    {\bf (a)}~Given a generic raster plot defined by a set of spikes occurring on specific addresses and at specific times, one may consider that this information consists of the repeated occurrence of groups of precise motifs of spikes that define ``polychronous groups'' (PGs). 
    {\bf (b)}~The generative model is defined by this set of motifs (here $4$ PGs) each defined by  synapses defined by weights (increasing with the radius of the black dots) with heterogeneous delay. 
    {\bf (c)}~Generalizing the core polychrony detection model (see Figure~\ref{fig:izhikevich}), one can define a layer of neurons that detect the identity and timing of these PGs~\citep{grimaldi_learning_2022}. % such that the addresses are the synapses of these neurons (which we therefore call here synapses).
    %Note that each afferent may be connected with multiple weights at different delays. The propagation of the afferent information through these delay may generate at each time step a synchronous motif on a subset of synapses. %These PG motifs were efficiently learned by a logistic regression.
    %{\bf (c)}~The output of the model provides with the predicted probability of occurrence of each PG motif at any time, which may be used to generate a spike as a Bernoulli trial, providing in this particular case with an exact identification of PGs occurrences.
    {\bf (d)}~Knowing the results of this detection, one may for illustration purposes highlight them by different colors in the raster plots, showing that in this synthetic examples, all spikes are now associated with a PG. 
     }
  \label{fig:THC} % TODO: citer biol cyb fig:THC
\end{figure}

Strikingly, thanks to the fact that a neuron can be involved in different polychronous groups, the number of coexisting polychronous groups far exceeds the number of neurons in the network, resulting in an unprecedented memory capacity of the system (see Figure~\ref{fig:THC}). In other neuronal models, an efficient use or detection of these spatio-temporal patterns embedded in the spike train comes with the integration of heterogeneous delays~\citep{guise_bayesian_2014,zhang_supervised_2020}. The recent ``multi-neuronal spike sequence detector'' architecture integrates the weight- and delay-adjustment methods by combining plasticity with the modulation of spike latency emission~\citep{susi_nmnsd-spiking_2021}. Additional models for detection of latency patterns are presented in the extensive (graph-centric) review on synchronization in time-varying networks~\citep{ghosh_synchronization_2021,ghosh_synchronized_2022}. This representation has potentially a much greater information capacity in comparison to other neural coding approaches through their connectivity and the possible coexistence of numerous superposed PGs~\citep{izhikevich_polychronous_2009}. Recently, by using a logistic regression model coupled with a temporal convolution, a model with heterogeneous delays was implemented to test the detection of the spiking motifs embedded in an event stream~\citep{grimaldi_learning_2022}. This allowed to detect a high number of superposed polychronous motifs in synthetic data, illustrating the computational benefit of such representations compared to that with a unique delay (see Figure~\ref{fig:THC}). As such, these models use the neural dynamics to handle input signals with different delays but do not explicitly take full advantage of the representation capacity offered by heterogeneous delays. 
%
%----------------------------%
%
% 
%----------------------------%
%%%%%%%%%%%%%%%%%%%%%%%%%%%%%%%%%%%%%%%%%%%%%
\subsection{Learning synaptic delays}
%%%%%%%%%%%%%%%%%%%%%%%%%%%%%%%%%%%%%%%%%%%%%
%%%%%%%%%%%%%%%%%%%%%%%%%%%%%%%%%%%%%%%%%%%%%
% Training delays ~\citep{state_training_2019}
First, the original model by Izhikevich use a simple STDP rule while a whole range of STDP-based learning rules may implement precise spiking motifs detection. For instance, to address how transmission delays and STDP can jointly determine these emergent pairwise activity-connectivity patterns, a recent study analyzed phase synchronization properties and coupling symmetry between two bidirectionally coupled neurons using both phase oscillator and conductance-based neuron models~\citep{madadi_asl_delay-dependent_2022}. Moreover, modified STDP rules have been used for synchronous coherence detection~\citep{perrinet_coherence_2002}, for the learning of specific receptive fields~\citep{perrinet_networks_2001}. They were also extended to recurrent Neuronal Networks~\citep{gilson_stdp_2010} or delay selection~\citep{datadien_right_2011}. In particular, this has been applied for recurrent networks of spiking neurons receiving oscillatory inputs~\citep{kerr_delay_2013} which targets for the selective potentiation of recurrent connections with different axonal and dendritic delays during oscillatory activity. More generally, our ability to track and respond to rapidly changing visual stimuli, such as a fast-moving tennis ball, indicates that the brain is capable of extrapolating the trajectory of a moving object to predict its current position, despite the delays that result from neural transmission. Specifically, the neural circuits underlying this ability can be learned through spike-timing-dependent synaptic plasticity and that these circuits emerge spontaneously and without supervision, demonstrating how the neural transmission delays can, in part, be compensated to implement the extrapolation mechanisms required to predict where a moving object is at the present moment~\citep{burkitt_predictive_2021}.

At the implementation level, a recent work proposed a bio-plausible unsupervised delay learning for extracting temporal features in spiking neural networks~\citep{nadafian_bio-plausible_2020}. The authors provided some mathematical proofs to show that their learning rule gives the ability to learn repeating spatio-temporal patterns. Applying this STDP-based rule on delays to the spiking neural network, experimental results were validated on a simple motion detection task, but were prone to convergence issues. Another model of synaptic delay-weight plasticity integrates synaptic delay plasticity into supervised learning and proposes a novel learning method that adjusts both the synaptic delays and weights of the learning neurons to make them fire precisely timed spikes~\citep{zhang_supervised_2020}. This was also presented by~\citep{wang_delay_2019} which propose a supervised delay learning algorithm for spiking neurons with temporal encoding, in which both the weight and delay of a synaptic connection can be adjusted to enhance the learning performance. Other models like that of~\citep{hazan_memory_2022} propose a weightless spiking neural networks that can perform a simple classification task which is applied to MNIST. In a recent paper~\citep{luo_supervised_2022}, authors propose a gradient descent-based learning algorithm for synaptic delays to enhance the sequential learning performance of a single spiking neuron. In this algorithm, information is encoded in the relative timing of individual neuronal spikes, and learning is performed based on the exact derivatives of the postsynaptic spike times with respect to presynaptic spike times.
In yet another computational model,~\citet{sun_learning_2016} show that the frequently activated polychronous neural groups can be learned efficiently by readout neurons with joint weight-delay spike-timing-dependent plasticity.

\subsection{Real-world applications}
%%%%%%%%%%%%%%%%%%%%%%%%%%%%%%%%%%%%%%%%%%%%%
%%%%%%%%%%%%%%%%%%%%%%%%%%%%%%%%%%%%%%%%%%%%%
%----------------------------%
% TODO
\begin{figure}%[t!]
  \centering
  \includegraphics[width=0.980\linewidth]{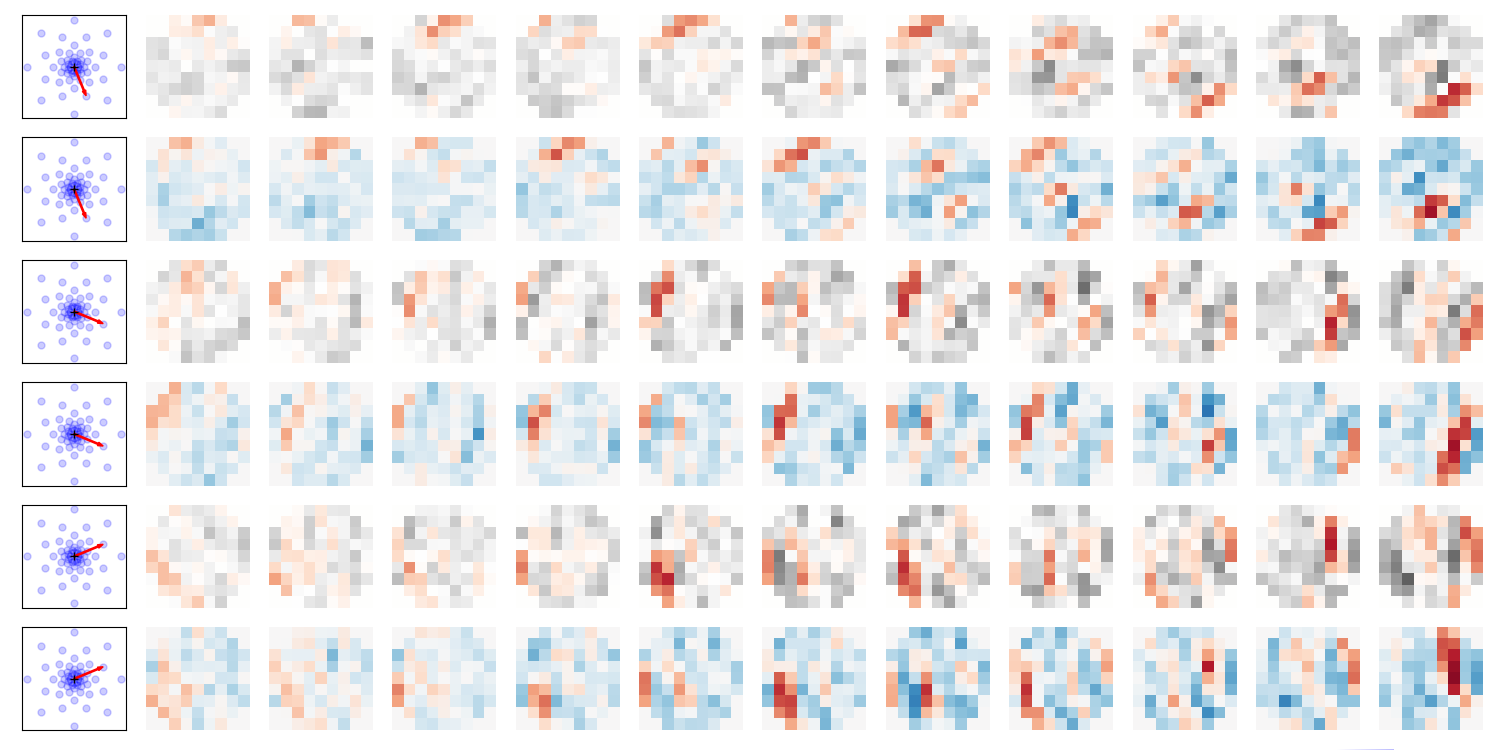}
    \caption{\textbf{Detecting visual motion in an event stream with heterogeneous delays.} Extending the polychrony detection model to the spatial domain,~\citet{grimaldi_learning_2022} have applied a supervised learning scheme to the detection of motion. The models parameters are represented by different spatio-temporal kernels, and we show three examples as pairs of rows, one targeting ON spikes, the other OFF spikes, the first column representing the corresponding motion detected. When trained on a set of natural images, it shows the emergence of localized, oriented kernels organized in a so-called push-pull organization for which weights to an ON spike are negatively proportional to that to an OFF cell~\citep{kremkow_push-pull_2016}. Global weight is globally decreasing from the lowest delay (right) to less recent information (left).
     }
  \label{fig:icip}
\end{figure}
% 
%----------------------------%

A second shortcoming of models derived from the polychronization model is their lack of applications in real-world scenarios. Indeed, most of these theoretical models are trying to reproduce neurobiological observations, while applications to machine learning methods such as image processing would further prove their plausibility. For instance, in a recent work,~\citet{ghosh_spatiotemporal_2019} propose a two-stage unsupervised-supervised system for the categorization of spatio-temporal actions from an event-based stream. The first stage learns spatio-temporal convolutional filters targeted to minimize event-removal related changes to a local spatio-temporal spike-event pattern. The second stage takes the output of the spatio-temporal filters as an input example containing multiple feature channels, and proceeds to train a classifier for recognition of spatio-temporal activity. For testing the system, two datasets are considered: DVS Gesture and a new action recognition dataset recorded for this work. Results demonstrate the ability of the system to outperform the state-of-the-art in event-based gesture recognition, along with demonstrating superior performance to other alternative ways of obtaining the first stage filters, thus showing the potential of such representation. 

There are more applications to image processing using spiking neural networks. For instance, a set of models are based on the design of micro-circuits with specific lateral interactions embedded with spatially anisotropic connections. Using this core computational unit, and extending it to computations on a topographic representation similar to that observed in the primary visual cortex of mammals, the anisotropic rules implemented a form of delayed activation. This result was based on a predictive model defined in the Bayesian framework (the so-called free-energy principle) which was able to account for temporal delays in the  system, both at the sensory and motor levels~\citep{perrinet_active_2014}, and in particular that ``the application of delay operators just means changing synaptic connection strengths to take different mixtures of generalized sensations and their prediction errors.'' Such a model was implemented at the network level and applied to various motion detection tasks. In essence, two neurons which were selective to specific motions were connected if the delay was coherent with the change in the position of their respective receptive fields~\citep{hogendoorn_predictive_2019}.  This was also implemented in a neural mass model, showing that such anisotropic connectivity may explain the emergence of tracking~\citep{khoei_motion-based_2013}, and further explored in a spiking neural network which reproduced the observation that neural activity was maintained during the trajectory of a smoothly moving dot even if it was momentarily blanked~\citep{kaplan_anisotropic_2013}. This led in particular to the proposal that such delay-based computations could explain diverse perceptual mechanisms, such as the so-called flash-lag illusion~\citep{khoei_flash-lag_2017}. However, these latter models used parametric rules for defining the weights. Extending such mechanisms with the ability of learning delays in a SNN will provide a breakthrough in the efficiency of these networks, and we will explore some exemplar results from neuromorphic engineering to get better insights on that aspect.
%
%%%%%%%%%%%%%%%%%%%%%%%%%%%%%%%%%%%%%%%%%%%%%
%%%%%%%%%%%%%%%%%%%%%%%%%%%%%%%%%%%%%%%%%%%%%
%%%%%%%%%%%%%%%%%%%%%%%%%%%%%%%%%%%%%%%%%%%%%
\section{Applications of precise spiking motifs in neuromorphic engineering}
\label{sec:neuromorphic}
%%%%%%%%%%%%%%%%%%%%%%%%%%%%%%%%%%%%%%%%%%%%%
%%%%%%%%%%%%%%%%%%%%%%%%%%%%%%%%%%%%%%%%%%%%%
%
Artificial intelligence has made huge advances in the past decades and deep learning algorithms, nowadays, outperform humans at complex tasks like natural image recognition or abstract strategy board games. Yet, machine learning algorithms suffer from adversarial attacks or a lack of generalization \change[Antoine]{capacity but}{capacity. But} their main weakness, compared to biological neural networks, rely on their poor energy efficiency. Neuromorphic engineering intends to mimic the neural bases of communication with a wide variety of technics, from strictly analog circuits to software-based neuromorphic systems, and to develop tools improving the capacities of current artificial intelligence~\citep{roy_towards_2019, javanshir_advancements_2022}. Because the reduced energy consumption of biological networks can be explained \add{in part} by the use of spikes and asynchronous responses to exchange information~\citep{maass_networks_1997}, neuromorphic chips use this parallel and event-based representation to \change{make}{perform} energy efficient computations. Another important distinction with classical von Neumann architectures is the localized memory of this new type of chips. It can be materialized by the capacity of the physical connections between the processing units to store information~\citep{markovic_physics_2020}. An example of such connection, directly inspired by synaptic plasticity, is the memristor~\citep{rasetto_challenges_2022} for which the resistance value can be dynamically adjusted. Using these event-based computations as a building brick, neuromorphic engineering proposes new hardware designs perfect to simulate SNNs and use the full power of asynchronous computations observed in biological systems. Even if some useful SNNs simulators run on GPUs~\citep{diesmann_nest_2003, hazan_bindsnet_2018, stimberg_brian_2019}, such event-based computing technics show their advantages in terms of frugality and rapidity only on neuromorphic chips. 

This field of research is inspired by neuroscientific advances and a computational formalism to design innovative architectures and, by artificially reproducing such mechanisms, it is interesting to study neural circuitry. Many connections can be drawn between neuromorphic engineering and computational neuroscience to aim at solving both research and technology challenges~\citep{zenke_visualizing_2021}. In this section, we make a description of the different neuromorphic hardware that have been developed and see how they can be used to deal with precise temporal motifs. 
\subsection{The emergence of novel computational architectures}
To our knowledge, the first neuromorphic circuit is the pulsed current-source synapse proposed by Carver Mead in 1989~\citep{mead_analog_1989}. It was implemented with transistors operating in the sub-threshold domain and responded to asynchronous events, but was not capable of discriminating two different spiking sequences with the same firing rate. Indeed, the postsynaptic membrane potential was increased by a step proportional to the input current but did not decrease in time, as it can be observed in biological neurons. Then, electronics circuits became more and more bio-realistic and, two decades later, ~\citep{bartolozzi_synaptic_2007} released the Diff-Pair Integrator (DPI) synapse that could reproduce the global dynamics of the biological neurons. The DPI circuit could multiplex in time spikes from different sources and became a potential  ``silicon coincidence detector''. Today, many devices are good candidates for implementing event-based algorithms and use the \change[Antoine]{AER}{address event representation (AER)}. They can be divided intro 3 major categories: digital, analog and mixed analog/digital platforms. For a more complete review the reader can refer to~\citep{schuman_survey_2017}, here we site the most popular ones. SpiNNaker~\citep{furber_overview_2013, furber_spinnaker_2020}, Loihi~\citep{davies_loihi_2018} and TrueNorth~\citep{merolla_million_2014} chips are widely used examples of digital hardware implementations. Compared to TrueNorth which exclusively implements a LIF neuron, SpiNNaker and Loihi offer some flexibility in terms of neuron model and allow for on-chip learning. This flexibility in the implementation comes at the cost of an increased energy consumption. Mixed-analog-digital systems were developed at the Stanford University: Neurogrid and Braindrop~\citep{benjamin_neurogrid_2014, neckar_braindrop_2019}. They are mostly used by computational neuroscientists to model brain activity with different levels of abstraction. BrainScaleS~\citep{schemmel_wafer-scale_2010} is another mixed-analog-digital system developed, just like SpiNNaker, for the Human Brain Project~\citep{markram_introducing_2011}. It is a wafer-scale neuromorphic hardware with analog components. Analog arrays (i.e. field programmable analog arrays (FPAA)) refer to the initial idea of neuromorphic hardware aiming at building strictly analog devices. The pulsed current source synapse and the DPI are examples of such devices, we can also mention the field programmable neural array~\citep{farquhar_field_2006} and the NeuroFPAA~\citep{cheng_fpaa_2009} specifically designed for neuromorphic systems. Due to their lack of generality and some issues specific to analog circuits, these fully analog devices are not yet widely used for neuromorphic computing. 
%
%----------------------------%
\begin{figure}
\centering
\includegraphics[width=0.980\linewidth]{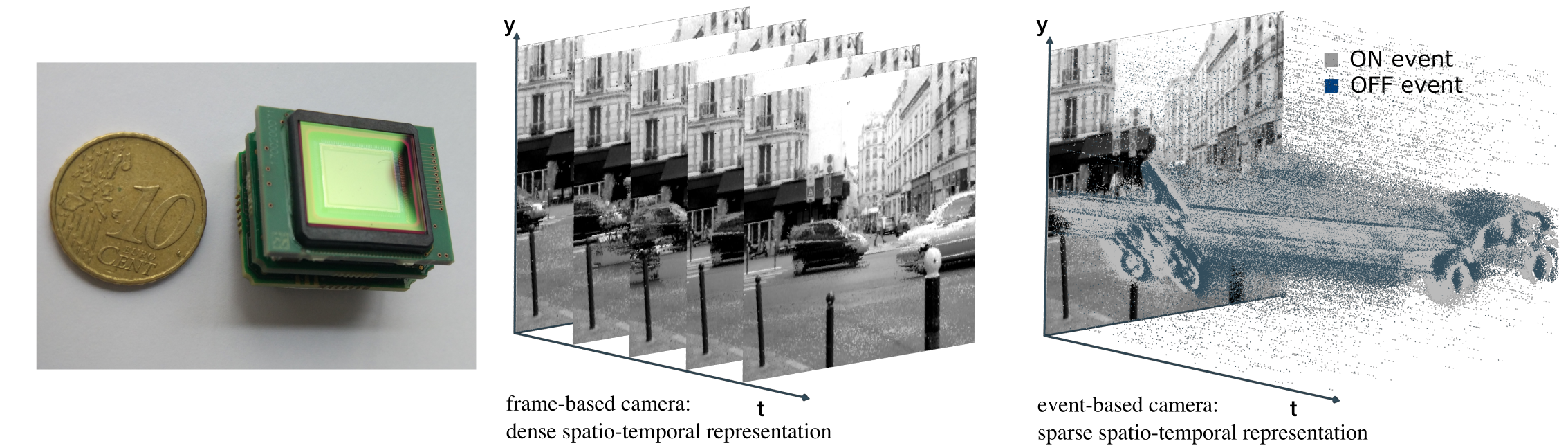}
\caption{\textbf{A miniature, event-based ATIS sensor.} Contrary to a classical frame-based camera for which a full dense image representation is given at discrete, regularly spaced timings, the event-based camera provides with events at the micro-second resolution. These are sparse as they represent luminance increments or decrements (ON and OFF events, respectively). Figure courtesy of Sio-Hoi Ieng (Sorbonne Université/UPMC, Institut de la Vision).}\label{fig:silicon_retina}
\end{figure}
%----------------------------%
%
Neuromorphic sensors are also developed with the idea to capture external stimuli more efficiently and closer to biological systems. A widely used example is the Dynamic Vision Sensors (DVS) which provide a stream of binary asynchronous events signaling detectable changes in luminance. These devices, also named ``silicon retinas'' (see Figure \ref{fig:silicon_retina}), show great improvements in terms of memory allocation, or power consumption, for the recording of a visual scene. We also report other event-based sensing devices for sound~\citep{chan_aer_2007} and touch~\citep{haessig_event-based_2020} but will focus on DVS for the next subsection about the use of dynamics embedded in event-based signals. 

%Using these event-based computations as a building brick, neuromorphic engineering proposes new hardware designs perfect to simulate SNNs and use the full power of asynchronous computations observed in biological systems. This field of research is inspired by neuroscientific advances and a computational formalism to design innovative architectures and, by reproducing such mechanisms, it is interesting to study neural circuitry. Many connections can be drawn between neuromorphic engineering and computational neuroscience to aim at solving both research and technology challenges~\citep{zenke_visualizing_2021}. Just as communication between neurons can be illustrated by a raster plot, information exchanges in neuromorphic hardware are described as event streams by an Address-Event Representation (AER). In this section, we will use indifferently \textit{events} or \textit{spikes} to describe the same binary and discrete processes and we will extend the review of the use of precise spike timing to this emerging field. There exist different types of event-based sensing devices for sound~\citep{chan_aer_2007} and touch~\citep{haessig_event-based_2020} and here we focus only on neuromorphic cameras to review different methods using efficiently the dynamics of event streams. Then, we dedicate one subsection to the implementation of delay-based algorithms on neuromorphic chips.  
%
\subsection{On the importance of spatio-temporal information in silicon retinas}
With the AER specification and their sub-millisecond temporal precision, DVS bring a new approach to the storage and processing of visual information. From their generative model, these sparse events are markers of the dynamics of the visual scene captured by the sensor. The dynamics of the event streams have to be used to make sense of the recorded information, and new algorithms are needed to solve efficiently classical computer vision tasks. 

In~\citep{lagorce_hots_2017}, \textit{time surfaces} are introduced as an event-driven 2D image of the delay between the last event recorded at the address of a pixel and the current time. An exponential decay is applied on this delay to obtain the analog values of the time surfaces. It gives more precision to represent recent events and offers an analogy with the LIF spiking neuron. Its a way to represent the local dynamics embedded in the event-based recordings, and with unsupervised learning on the event stream based on the cosine similarity they can capture repeating motifs within the input signals. Learned time surfaces can be used for object recognition~\citep{lagorce_hots_2017, sironi_hats_2018, maro_event-based_2020, grimaldi_homeostatic_2021,grimaldi_robust_2022} and show that this method could be efficiently applied to state-of-the-art benchmarks.

More generally, three-dimensional convolutions in both space and time are another representation of the spiking motifs embedded in the event stream~\citep{ghosh_spatiotemporal_2019, grimaldi_learning_2022, yu_stsc-snn_2022}. With their additional temporal dimension, their kernels can capture multiple events on the same pixel address, as long as they belong to the local temporal window (see Figure~\ref{fig:icip}). This representation is only limited by the time step used for the discretization of the signal, this factor defines the temporal precision of the representation. Other methods make direct use of the precise timing of events captured by the DVS to solve optical flow and time-to-contact challenges~\citep{benosman_event-based_2014, clady_asynchronous_2014, tschechne_bio-inspired_2014}, inferring depth~\citep{hidalgo-carrio_learning_2020}, feature detection and tracking~\citep{dardelet_event-by-event_2021}, motion segmentation~\citep{stoffregen_event-based_2019} or the simultaneous localization and mapping problem~\citep{kim_real-time_2016}. This non-exhaustive list of complex task solving is not directly linked to biological processes, but show the potential of the precise temporal resolution of neuromorphic retina-like cameras. By essence, these sensors offer a novel view of visual information processing due to the asynchronous responses of the different pixels. With this type of signal, the use of spatio-temporal motifs embedded in the event streams are essential to solve high level visual tasks. 
\subsection{Computations with delays in neuromorphic hardware}
For the rest of this section, we report examples of implementations of event-based algorithms using precise spatio-temporal motifs on neuromorphic hardware. 
\citep{hussain_deltron_2012} implemented a delay-learning algorithm on an analog chip. Online learning on neuromorphic chips is still a challenge today and for this work, only the detection of spiking motifs was done on the analog architecture, training was performed digitally and based on the tempotron learning rule~\citep{gutig_tempotron_2006}. In addition to delay learning, a group at the University of West Sydney developed a neuromorphic implementation of multiple synaptic plasticity learning rules~\citep{wang_neuromorphic_2015}. They show that STDP and Spike Timing Dependent Delay Plasticity rules could be implemented in both a digital and an analog chip. From the network parameters and the physical limitations to store it on-chip, they prove that the digital implementation is way easier to scale up and that an external memory would be needed for a larger network. The same group presented a FPGA hardware implementation of polychronous networks in which propagation delays are learned in a supervised manner, based on the expected firing time of the post-synaptic neuron~\citep{wang_fpga_2014}. %This paper proposes a supervised delay learning algorithm for spiking neurons with temporal encoding, in which both the weight and delay of a synaptic connection can be adjusted to enhance the learning performance. 
\citet{pfeil_neuromorphic_2013} implemented STDP on a mixed-analog-digital chip to simulate the sound localization processes observed on the barn owl auditory system~\citep{gerstner_neuronal_1996}. Coherence detection on input spikes coming from two sources was obtained with a 50 $ns$ precision. They claim that this unsupervised learning denoises the input and compensates for variations between neural components. The variability of response of the analog components can be compensated by population coding for responses robust to noise and this phenomenon is also observed in biological neural networks~\citep{boerlin_spike-based_2011}. 
A recent work makes implementation of a sparse Vector Symbol Architecture binding operation on the Loihi chip, delay lines and coincidence detection are used to compute the binding operation~\citep{renner_sparse_2022}. They highlight the fact that using delays can be expensive \add[Laurent]{notably in memory bandwidth} because incoming spikes have to be stored in blocks with a temporal dimension equivalent to the maximal delay. Note that this problem is also due to the algorithm used in this paper and that the analog chips must not suffer from this type of limitation. 

Online on-chip learning and computations with delays are still emerging in neuromorphic engineering. The technical challenges linked to the development of this type of implementation and the growing interest in delay learning make advances in this field interesting for the future of computations with precise spatio-temporal motifs. While improvements are still to be made, neuromorphic chips seem to be a good candidate to efficiently make use of these particular features. 
%
%they made a Spike-Timing Neuromorphic Architecture with a dendritic arbor to implement the delays and a transfer function for the weights which is well-suited for areas such as pattern recognition and natural language processing. In this paper, we formally describe the STPU, implement the STPU on a field programmable gate array, and show measured performance data.\citep{hill_spike-timing_2017} not accessible
%
%review on neuromorphic computations for robotic\citep{sandamirskaya_neuromorphic_2022} (not accessible for the moment)
%
%We apply delay-based reservoir computing as the information processing core, along with a novel training and labelling method. Different from the conventional ECG classification techniques, this computation model is a end-to-end dynamic system that mimics the real-time signal flow in neuromorphic hardware~\citep{liang_neuromorphic_2022} 
%
%Real-time cortical simulation on neuromorphic hardware -> https://royalsocietypublishing.org/doi/full/10.1098/rsta.2019.0160
%
% solutions usuellement utilisées - surrogate gradient
%for other section: A hypothesis for temporal coding of young and mature granule cells(https://www.frontiersin.org/articles/10.3389/fnins.2013.00075/full)
%
%%%%%%%%%%%%%%%%%%%%%%%%%%%%%%%%%%%%%%%%%%%%%
%%%%%%%%%%%%%%%%%%%%%%%%%%%%%%%%%%%%%%%%%%%%%
%%%%%%%%%%%%%%%%%%%%%%%%%%%%%%%%%%%%%%%%%%%%%
\section{Discussion}\label{sec:discussion}
%%%%%%%%%%%%%%%%%%%%%%%%%%%%%%%%%%%%%%%%%%%%%
%%%%%%%%%%%%%%%%%%%%%%%%%%%%%%%%%%%%%%%%%%%%%
%%%%%%%%%%%%%%%%%%%%%%%%%%%%%%%%%%%%%%%%%%%%%
\subsection{Summary}
%%%%%%%%%%%%%%%%%%%%%%%%%%%%%%%%%%%%%%%%%%%%%
%%%%%%%%%%%%%%%%%%%%%%%%%%%%%%%%%%%%%%%%%%%%%
%: >>>>>>>> Laurent is here <<<<<<<<<<<<<<
In this review paper, we have presented recent evidence for the role of precise spiking motifs in neuroscience. In particular, we have shown that such particular motifs may play a crucial role in neurobiology, that they may be understood at the theoretical and computational levels and that they may have numerous applications in neuromorphic engineering. In particular, we have shown that:
\begin{itemize}
  \item  the efficiency of neural systems, and in particular the visual system, imposes strong constraints on the structure of neural activity which highlights the importance of precise spike times,
  \item  growing evidence from neurobiology proves that neural systems are more than integrators and may use synchrony detection in different forms: synfire chains, travelling waves on arbitrary spiking motifs, \add[Laurent]{and notably that a an encoding based on precise spiking motifs may provide huge computational benefits},
  \item  many theoretical models already exist taking into account the specificity of spiking motifs, notably by using heterogeneous delays,
  \item  using precise spiking motifs could ultimately be a key ingredient in neuromorphic systems to reach similar efficiencies as biological neural systems.
\end{itemize}
 Overall, our reviewing effort has shown that a growing community is focusing on that aspect. This community is based on solid and validated evidence which is getting novel grounds thanks to current technical advances. Moreover, we have also shown that this community is highly diverse, operating in biology, computational neuroscience or neuromorphic engineering. As a consequence, the global effort is still largely scattered, which limits its larger acceptance in neuroscience. 

\subsection{Limits}
Additionally, the different models of spike motif detection and learning that we have presented at these different levels (neurobiological, theoretical, neuromorphic) individually present limitations that prevent their widespread application in neuroscience.

First, many models are based on a discretization of time. This assumption is important to allow for a useful representation of neuronal information in order to be processed in computers. This treatment amounts to transforming spike trains into a matrix form for processing in classical machine learning algorithms. This assumption therefore implies an ineffective use of the memory, as this representation transforms the sparse representation of a spike sequence into dense matrices. In addition, this representation can induce errors due to the discretization and the scale of temporal sampling. Finally, this representation encourages the use of classical methods, which are not adapted to disruptive applications such as event-driven representations.

Moreover, the learning of patterns is often done in a supervised way. Indeed, the problem of detecting polychronous groups implies to infer both the address and the precise time of occurrence of these motifs. Most of the models we have presented are based on the assumption that at least one of its variables is known: either the pattern, its identity, or its time of occurrence. This constraint is to be put in parallel with the way a biological nervous system works in which learning is done autonomously, i.e. without supervision. However, we can note that some models can perform such learning, but only in the case of data for which the motifs are easily separable. More generally, to reproduce the efficiency of biological systems, one should account for the different temporal scales of adaptation, from seconds to years. For instance, the scaffolding of neural assemblies seems to follow critical periods during development~\citep{dard_rapid_2022}.

A final limitation of the models we have presented is that they consist of a single processing layer that links an input to an output. However, we have seen that the neurobiological system uses processing loops within hierarchical graphs. In general, these systems are bidirectionally connected across different layers (for instance cortical areas), but also within a layer, as was for instance used by~\citet{izhikevich_polychronization_2006}. The whole system forms a dynamical model which may be considered globally during the learning phase, yet while taking account the constraints of the system, for instance the lack of a global clock, or the cost of fully-connected topologies. Moreover,  these processes have to be distinguished from judgements on timing such as temporal order processing (judging when one event happens relative to another) or duration estimation (measuring how long an event lasts)~\citep{coull_distinction_2022}.
\subsection{Perspectives}
The limits that we have presented can be treated individually in each model, as evidenced by individual efforts which try to overcome them. However, to propose a real breakthrough, we believe that future venues should provide with a unified, interdisciplinary approach, with applications to real-world, ecological scenarios and with open and reproducible methods. 

First, as we already noted, the effort is still largely scattered. This is in part due to the fact that interaction between neurobiology, theoretical and computational neuroscience and neuromorphic engineering are still scarce as of today. It was largely demonstrated that close, bidirectional interactions are essential to foster breakthroughs. For instance, the design of model-driven protocols has proven to be essential in modern neuroscience. Also, if neural networks were essential in shaping modern-day machine learning, e-g. computer vision using deep-learning, spiking neural networks should prove essential in future emerging technologies.

In that perspective, it is essential that such models are tested on ecologically-relevant, real-world scenarios. Indeed, classic convolutional neural networks emerged as optimal solutions for instance to classify static images into categories, yet they are not well adapted for processing dynamic, multimodal sensory flows. The emerging necessity to be able to process more complex flows, such as the multiple flows of information in a car designed for autonomous driving, necessitates to modify such modelling paradigms, and in particular, to take into account that the generated actions may modify the sensory inputs. Notably, the protocols used as well in neurobiology, theory or engineering should take into account this novel levels of complexity.

Ultimately, the community should encourage the adoption of open, reproducible science. Indeed, the different models that we have displayed often come with the tools necessary to reproduce the results obtained. This is true in neurobiology~\citep{bellec_fitting_2021}, in theoretical neuroscience~\citep{izhikevich_polychronization_2006} or in engineering~\citep{grimaldi_robust_2022}. This aspect is essential to foster the emergence of interdisciplinary projects, such as model-driven neurobiological experiments or biologically-inspired neuromorphic engineering. Solutions exist to optimize these collaborations~\citep{panahi_generative_2021} and suggest the emergence of a novel paradigm for scientific advances in neuroscience~\citep{tolle_fourth_2011} i.e., by using data exploration in which the scientific models are fit to the data by learning algorithms. As such, this review aims at paving the way to openly share the variety of resources and to offer a unified view on the role of precise spiking motifs in neuroscience.
%%%%%%%%%%%%%%%%%%%%%%%%%%%%%%%%%%%%%%%%%%
%%%%%%%%%%%%%%%%%%%%%%%%%%%%%%%%%%%%%%%%%%
\vspace{6pt} 
%%%%%%%%%%%%%%%%%%%%%%%%%%%%%%%%%%%%%%%%%%
%% optional
%\supplementary{The following supporting information can be downloaded at:  \linksupplementary{s1}, Figure S1: title; Table S1: title; Video S1: title.}


%%%%%%%%%%%%%%%%%%%%%%%%%%%%%%%%%%%%%%%%%%
% \authorcontributions{Conceptualization, LP; writing---original draft preparation, all authors; writing---review and editing, all authors; supervision, LP; funding acquisition, LP and JM. All authors have read and agreed to the published version of the manuscript.}

\funding{
This research was funded by the European Union ERA-NET CHIST-ERA 2018 research and innovation program under grant number ANR-19-CHR3-0008-03 (``\href{http://aprovis3d.eu/}{APROVIS3D}''). % Agence Nationale de la Recherche %HL and 
LP, CB and AG received funding from the ANR grant number ANR-20-CE23-0021 ``\href{https://laurentperrinet.github.io/grant/anr-anr/}{AgileNeuroBot}''. %
% LP received funding 
LP received support from the french government under the France 2030 investment plan, as part of the Initiative d’Excellence d’Aix-Marseille Université – A*MIDEX number AMX-21-RID-025 ``\href{https://laurentperrinet.github.io/grant/polychronies/}{Polychronies}''.}

\dataavailability{This work is made reproducible using the following tools. First the manuscript's source code, and the code reproducing all figures are available on \href{https://github.com/SpikeAI/XXX}{GitHub}. %~\citep{Grimaldi2023review}. 
The paper is available as an \href{https://arxiv.org/abs/YYYY.XXX}{arXiv preprint} with links to previous versions. Find the associated \href{https://www.zotero.org/groups/4562620/polychronies}{Zotero group} which was used to regroup relevant literature on the subject.} 

\acknowledgments{We would like to thank Hugo Ladret and Frédéric Chavane for valuable feedback while writing this review. Many thanks to Simon Thorpe for the useful discussions and references on the latency of the visual system in human and non-human primates.}

\conflictsofinterest{The authors declare no conflict of interest. The funders had no role in the design of the study; in the collection, analyses, or interpretation of data; in the writing of the manuscript; or in the decision to publish the results.} 

\begin{adjustwidth}{-\extralength}{0cm}

%%%%%%%%% REFERENCES
\reftitle{References}

%\bibliographystyle{Definitions/chicago2}
\bibliography{references}

%\begin{thebibliography}{00}
%\printbibliography[heading=none]
%\end{thebibliography}

% For the MDPI journals use author-date citation, please follow the formatting guidelines on http://www.mdpi.com/authors/references
% To cite two works by the same author:~\citeauthor{ref-journal-1a} (\citeyear{ref-journal-1a},~\citeyear{ref-journal-1b}). This produces: Whittaker (1967, 1975)
% To cite two works by the same author with specific pages:~\citeauthor{ref-journal-3a} (\citeyear{ref-journal-3a}, p. 328;~\citeyear{ref-journal-3b}, p.475). This produces: Wong (1999, p. 328; 2000, p. 475)

\end{adjustwidth}
\end{document}